# Unconventional Unidirectional Magnetoresistance in vdW Heterostructures

I-Hsuan Kao[1], Junyu Tang[2], Gabriel Calderon Ortiz[3], Menglin Zhu[3], Sean Yuan[1], Rahul Rao[4], Jiahan Li[5], James H. Edgar[5], Jiaqiang Yan[6,7], David G. Mandrus[6,7], Kenji Watanabe[8], Takashi Taniguchi[9], Jinwoo Hwang[3], Ran Cheng[2,10,11], Jyoti Katoch[1], and Simranjeet Singh[1,*]

[1]Department of Physics, Carnegie Mellon University, Pittsburgh, PA, 15213, USA

[2]Department of Physics and Astronomy, University of California, Riverside, CA 92521, USA

[3]Department of Materials Science and Engineering, The Ohio State University, Columbus, OH 43210, USA

[4]Materials and Manufacturing Directorate, Air Force Research Laboratory, Wright-Patterson Air Force Base, Dayton, OH, 45433, USA

[5]Tim Taylor Department of Chemical Engineering, Kansas State University, Manhattan, Kansas 66506, USA

[6]Materials Science and Technology Division, Oak Ridge National Laboratory, Oak Ridge, Tennessee 37831, USA

[7]Department of Materials Science and Engineering, The University of Tennessee, Knoxville, TN 37996, USA

[8]Research Center for Electronic and Optical Materials,

National Institute for Materials Science, 1-1 Namiki, Tsukuba 305-0044, Japan

[9]Research Center for Materials Nanoarchitectonics,

National Institute for Materials Science, 1-1 Namiki, Tsukuba 305-0044, Japan

[10]Department of Electrical and Computer Engineering, University of California, Riverside, CA 92521, USA

[11]Department of Material Science and Engineering, University of California, Riverside, CA 92521, USA

---

[*]Email: simranjs@andrew.cmu.edu

**Electrical readout of magnetic states is a key to realize novel spintronics devices for efficient computing and data storage. Unidirectional magnetoresistance (UMR) in bilayer systems, consisting of a spin-source material and a magnetic layer, refers to a change in the longitudinal resistance upon the reversal of magnetization, which typically originates from the interaction of spin-current and magnetization at the interface[1-9]. Because of UMR's linear dependence on applied charge current and magnetization, it can be used to electrically read the magnetization state. However, in conventional spin-source materials, the spin-polarization of an electric-field induced spin current is restricted to be in the film plane[10-14] and hence the ensuing UMR can only respond to the in-plane component of the magnetization. On the other hand, magnets with perpendicular magnetic anisotropy (PMA) are highly desired for magnetic memory and spin-logic devices[14,15], while the electrical read-out of PMA magnets through UMR is critically missing. Recently, topological semimetals with low-crystal symmetries, such as $WTe_2$, emerged as a unique spin-source material platform, wherein out-of-plane spin polarization can be efficiently generated through spin-galvanic effects[16,17], which in principle can enable electrical read-out of PMA magnets through UMR. Here, we report the discovery of an unconventional UMR in bilayer heterostructures of a topological semimetal ($WTe_2$) and a PMA ferromagnetic insulator ($Cr_2Ge_2Te_6$, CGT), which allows to electrically read the up and down magnetic states of the CGT layer by measuring the longitudinal resistance. Our theoretical calculations based on a tight-binding model show that the unconventional UMR originates from the interplay of crystal symmetry-breaking in $WTe_2$ and magnetic exchange interaction across the $WTe_2$/CGT interface. Combining with the ability of $WTe_2$ to obtain magnetic field free switching of the PMA magnets[17], our discoveries open an exciting pathway to achieve two-terminal magnetic memory devices that operate solely on the spin-orbit torque and UMR[11,14], which is critical for developing next-generation non-volatile and low-power consumption data storage technologies.**

Since the discovery of the giant magnetoresistance and its tunneling counterpart[18-21], which led to the development of magnetic data storage devices, other magnetoresistance effects have been experimentally observed in magnetic heterostructures and has been employed to electrically read magnetic states[3,22-27]. Among other effects, unidirectional magnetoresistance (UMR) refers to a change in the longitudinal resistance of a bilayer system, consisting of a spin-source material and a magnetic layer, upon the magnetization reversal, which originates from the interaction of spin-current in the spin-source material and the magnetization in the adjacent magnetic layer[1-9]. A change in this magnetoresistance, $R_{UMR}$, phenomenologically obeys: $R_{UMR} \propto (\boldsymbol{m} \cdot \boldsymbol{\sigma})J$, where $J$ is the applied charge current density, $\boldsymbol{\sigma}$ is the spin polarization generated by the spin-galvanic effect in the spin-source material[22,28,29], and $\boldsymbol{m}$ specifies the magnetization direction of the magnetic layer. Distinct from other forms of magnetoresistance, UMR features linear dependence on both $J$ and $\boldsymbol{m}$, thus it can be used to monitor the magnetic state of a magnetic system; and the UMR is finite only when the spin polarization has a component collinear with the magnetization, i.e., $\boldsymbol{m} \cdot \boldsymbol{\sigma} \neq 0$. In conventional spin-source materials such as heavy metals and topological insulators, a charge current applied in the film plane (e.g., $J \parallel \hat{\boldsymbol{x}}$) can only generate a spin accumulation polarized in-plane and transverse to the charge current, i.e., $\boldsymbol{\sigma} \parallel \hat{\boldsymbol{y}}$ [10-13]. Consequently, the UMR response in prevailing bilayer systems is only limited to the form of $R_{UMR} \propto (\boldsymbol{m} \cdot \hat{\boldsymbol{y}})J \propto$ $(m_y)J$, and thus cannot be employed to read the up and down states of a ferromagnet (FM) with PMA. Nonetheless, FMs with PMA are highly desired for magnetic memory and spin-logic devices to realize ultra-fast operation, thermally stable nanometer sized magnetic-bits, and achieving attojoule-class logic gates. In this regard, the key to achieve a non-volatile writing operation is to utilize the spin-orbit torque (SOT) driven magnetic switching, wherein an out-of-plane magnetization can be efficiently manipulated by the SOT originating from the spin current generated in a nearby spin-source layer[10-14]. However, SOT-based planar magnetic memory devices considered so far all rely on a three-terminal configuration[11,14], wherein a MTJ is integrated on top of spin-source material to read the magnetic state through tunnel-magnetoresistance effect, which critically affect the footprint of the individual memory nodes. At present, an SOT-based two-terminal magnetic memory device, like commercially available MTJ-based magnetic memory devices, is critically missing due to the lack of identified magnetoresistance phenomena to distinguish the up and down magnetic states of a PMA magnet.

Here, we report the discovery of an unconventional UMR in bilayer heterostructures consisting of a topological semimetal (WTe$_2$)[30,31] and a ferromagnetic insulator (Cr$_2$Ge$_2$Te$_6$, CGT)[32-34], which allows us to directly read the up and down states of the PMA magnetic layer through longitudinal resistance measurements. The UMR originates from the interplay of crystal symmetry-breaking in WTe$_2$ and magnetic interactions across the WTe$_2$/CGT interface. WTe$_2$ is believed to be a type-II Weyl semimetal[30,35] whose T$_d$ phase has a crystal structure with broken mirror symmetry in the *ac*-plane (Fig. 1a), which allows for the generation of spin current with a substantial out-of-plane spin polarization ($\sigma_z$) driven by a charge current applied along the crystallographic *a*-axis (left panel of Fig. 1b) owing to the spin-galvanic effect[16,17,36-38]. The interactions between $\sigma_z$ in WTe$_2$ and the magnetization in CGT gives rise to a UMR (Fig. 1b right) that can be used to distinguish between the up and down magnetic states through longitudinal resistance measurements, i.e., lower (higher) resistance when **m** and $\sigma_z$ are parallel (antiparallel). In the following, we first present our main experimental results and then provide a theoretical interpretation basing on a minimal tight-binding model calculation, shedding light on the microscopic origin of the unconventional UMR in WTe$_2$/CGT bilayer systems.

For our experiments, we employ atomically thin flakes of WTe$_2$ (few-layers) and CGT (10-15 nm), and techniques of mechanical dry transfer and standard device fabrication are used to assemble van der Waals (vdW) heterostructures and devices of WTe$_2$/CGT bilayer[17] (see Methods and Supplementary note 1). In the main text, we present experimental data from three samples, i.e., namely device A and device B with 2-layers and 3-layers of WTe$_2$ respectively, and a control sample (device C). The optical micrograph of device A is shown in Fig. 1c (lower panel) along with a side-view schematic of a typical device (top panel). The flake thicknesses are confirmed by a combination of optical contrast and atomic force microscopy (AFM). The bilayer WTe$_2$ flake used in device A is shown in the inset to Fig.1c, wherein the thickness is confirmed with AFM (Fig. 1f) and crystallographic *a*-axis and *b*-axis are identified using polarized Raman spectroscopy (Fig. 1d)[39]. Raman spectra are collected by rotating the polarization of the incident laser for different angles relative to the current electrode direction and the normalized intensity at Raman peak 163 cm$^{-1}$ at each polarization angle is plotted as the polar plot shown in Fig. 1d. The maximal intensity around 0 degree indicates the *a*-axis of WTe$_2$ in device A is aligned to the current electrodes. We have also carried out atomic structure characterization of our measured devices using scanning transmission electron microscopy (STEM). A STEM cross-sectional image of a measured device (device B), consisting of tri-layer of WTe$_2$ and 13.2 nm CGT, is shown in Fig.

1e, wherein an atomically clean and sharp WTe$_2$/ CGT interface is clearly visible. For all measured devices, the crystallographic orientation of WTe$_2$ is confirmed by polarized Raman spectroscopy and STEM (see Supplementary Note 2). Also, at low temperature the measured resistance of individual CGT flakes is orders of magnitude larger as compared with the resistance of the WTe$_2$/CGT bilayer, as confirmed in a control device (Fig. 1g and see Supplementary Note 3), indicating that most of the applied charge current to the WTe$_2$/CGT bilayer flows through the WTe$_2$ layer.

In our bilayer devices, consisting of an atomically thin WTe$_2$ layer and an insulating magnet, the electric transport in the WTe$_2$ layer is tunable by applying an electrostatic gate voltage, which adjusts the electron chemical potential in the WTe$_2$ layer. The measured longitudinal resistance ($R_{xx}$) for device A as a function of top gate voltage ($V_g$) at 10 K is shown in Fig. 2a, wherein $R_{xx}$ increases with increasing $V_g$, indicating that the bilayer WTe$_2$ is likely hole-doped due to the charge transfer across the interface. The hysteresis loop in $R_{xx}(V_g)$ is also indicative of ferroelectric behavior in WTe$_2$ as previously reported[40,41]. First, we discuss the experimental observation of anomalous Hall effect (AHE) in WTe$_2$/CGT system. The Fig. 2b shows the measured Hall resistance ($R_{yx}$) as a function of out-of-plane magnetic field ($B_z$) in device A at different gate voltages. The observe saturation in $R_{yx}$ at $\pm B_z$ agrees well with magnetic characteristics of CGT flakes, i.e., magnetization saturation around $B_z \sim 50\ mT$ with negligible coercivity[33,34]. By defining $R_{yx} = \Delta R_{AHE} m_z$, where $\Delta R_{AHE}$ is the change in measured AHE resistance signal, we can extract $\Delta R_{AHE}$ as a function of gate voltage and is shown in Fig. 2c. We observe that the magnitude of $\Delta R_{AHE}$ is strongly suppressed (enhanced) by positive (negative) gate voltages. Next, we perform AHE hysteresis loop measurement at $V_g = 0$ for various temperatures and the obtained $\Delta R_{AHE}$ as a function of temperature as shown in Fig. 2d, wherein the AHE signal vanishes near the Curie temperature of CGT (~60 K) showing that observed AHE is due to magnetism in CGT[32,42] (see Supplementary Note 7). The upper panel of Fig. 2e shows the measured AHE hysteresis loop in device B with positive and negative current applied to confirm that $\Delta R_{AHE}$ does not switch sign with current polarity reversal as expected for a linear current response. To eliminate thermal contribution irrelevant to AHE, we also extract the symmetric (antisymmetric) component of the Hall resistance, with respect to the current polarity, using $R_{yx}^{symm(asym)} = \frac{R_{yx}(+I_{DC}) \pm R_{yx}(-I_{DC})}{2}$ and as expected observe a negligible antisymmetric component of the measured AHE response (lower panel of Fig. 2e). Like device A, the device B also shows a hole-doped transport and ferroelectric like loop (inset to Fig. 2f), albeit a lower channel resistance due to 3-layers of WTe$_2$, and a gate voltage dependent $\Delta R_{AHE}$ (Fig. 3f), i.e., suppression of AHE at positive $V_g$, is observed. AHE can originate from either the phenomenon associated with spin-galvanic effects[24,43,44], i.e., spin Hall effect driven spin current in WTe$_2$ and its magnetization dependent scattering at the interface, or due to the magnetic proximity effects in WTe$_2$ [27,45-48], e.g., the Berry curvature induced by an exchange interaction from an out-of-plane magnetization of CGT.

Next, we present results showing the experimental realization of unconventional UMR in WTe$_2$/ CGT bilayers to electrically read the out-of-plane magnetization of CGT. As shown in Fig. 3a, when a positive charge current is applied along the crystallographic *a*-axis of WTe$_2$ in device A, there is a clear step in resistance, proportional to $m_z$, in the measured raw longitudinal resistance ($R_{xx}^{raw}$) as a function of $B_z$, indicating the presence of unconventional UMR. By considering the saturated branches at high-field (>100mT), we model the measured longitudinal resistance by considering $R_{xx}^{raw} = \Delta R_{UMR} m_z + R_{xx,\alpha} B_z^2 + R_{xx,\beta} B_z + R_0$, where $\Delta R_{UMR}$ is the UMR resistance, $R_{xx,\alpha}$ ($R_{xx,\beta}$) is parameter for quadratic (linear) $B_z$ dependence, and $R_0$ is the field independent part of the longitudinal resistance. The quadratic $B_z$ dependence is due to the ordinary magnetoresistance[31] while the slight linear $B_z$ dependence correction is due to the contribution of ordinary Hall effect (OHE) resulting from the non-uniform width of the flakes in our devices. Fig. 3b shows the measured longitudinal resistance $R_{xx}$, after excluding effects other than UMR, with positive and negative current applied, where $\Delta R_{UMR}$ switches sign with current polarity reversal, as expected for a quadratic current response. We extract the symmetric (antisymmetric) part of the $R_{xx}$, with respect to the current polarity, as follows $R_{xx}^{symm(asym)} = \frac{R_{xx}(+I_{DC}) \pm R_{xx}(-I_{DC})}{2}$. As expected, UMR response, due to its unidirectional nature, is negligible in the symmetric part of $R_{xx}$. The small peaks in $R_{xx}^{symm}$ is likely due to the linear magnetoresistance effect proportional to $(\boldsymbol{m} \cdot \boldsymbol{\sigma})^2$, such as spin-galvanic magnetoresistance[23]. Next, we focus on UMR by showing $R_{xx}^{asym}$ as a function of $B_z$ in device A at various gate voltages (Fig. 3c). We extract $\Delta R_{UMR}$ as a function of $V_g$ and is shown in Fig. 3d, which clearly shows that $\Delta R_{UMR}$ is enhanced (suppressed) at positive (negative) $V_g$. Next, UMR measurements are performed at various temperatures (details in Supplementary note 7) to obtain the $\Delta R_{UMR}$ as a function of temperature (Fig. 3e). As expected, $\Delta R_{UMR}$ reduces as the temperature approaches Curie temperature of CGT, indicating that the d UMR is indeed due to the magnetic coupling between WTe$_2$ and CGT magnetization. Next, we examine the current dependence of UMR (Fig. 3f) by extracting the $\Delta R_{UMR}^{+/-}$ at various current magnitudes, where $\Delta R_{UMR}^{+/-}$ is the $\Delta R_{UMR}$ measured at positive (negative) current (details in Supplementary note 6). The magnitude of UMR scales with current in the low current region, while at higher current region it shows a deviation from linear behavior which is likely due to increase of temperature in CGT caused by significant Joule heating. For device B, with 3-layers of WTe$_2$, when a charge current is applied along the *a*-axis, the $R_{xx}^{raw}$ as a function of $B_z$ is shown in Fig. 3g. The extracted $R_{xx}^{asym/symm}$ in device B (Fig. 4h) shows UMR results, which are consistent with UMR data obtained using device A. Finally, the $\Delta R_{UMR}$ in device B also shows a gate voltage dependence (like device A), wherein the magnitude of UMR is suppressed by negative gate voltage. Note the $\Delta R_{UMR}$ is negative in device B, because the UMR can gain a sign change depends on whether the positive current direction is defined along the $+\hat{a}$ or $-\hat{a}$ of WTe$_2$. Unfortunately, there is no non-invasive probe to explicitly distinguish $+\hat{a}$ and $-\hat{a}$ of WTe$_2$, therefore there can be a 180° rotation about the *c*-axis of WTe$_2$ depending on which end of WTe$_2$ is connected to the ground. The UMR will switch sign because the sign of $\sigma_z$ is determined by whether the charge current is applied along $+\hat{a}$ and $-\hat{a}$ in WTe$_2$.

On the other hand, magnetoresistance behavior is expected to be completely different, i.e., UMR due to out-of-plane magnetization should vanish due to the absence of $\sigma_z$ when the current is applied along the *b*-axis. To examine this, we prepare a control device C (Fig. 4a), consisting of a 3-layers of WTe$_2$ and 8.1 nm CGT, where the *b*-axis is aligned to the current electrodes, as confirmed by polarized Raman spectroscopy (Fig. 4b). As presented in Fig. 4c, the $R_{xx}(V_g)$ in device C shows hole-doped and ferroelectric behavior (like device A and device B). Measured $R_{yx}$ as a function of $B_z$ in device C at various gate voltages is shown in Fig. 4d depicting a clear AHE, indicating that AHE is independent of whether current is applied along the *a*- or *b*-axis. On the other hand, in a striking contrast, $R_{xx}^{asym}$ as a function of $B_z$ at various gate voltages measured in device C (Fig. 4e) shows no signature of UMR response when the current is applied along the *b*-axis. To better compare the UMR response in WTe$_2$/ CGT devices, we plot the normalized UMR, i.e., $|\Delta R_{UMR}/R_0|$ with $R_0$ as the magnetic field independent part of the longitudinal resistance, as a function of charge current density $|J_{DC}|$ and results from three measured devices is shown in Fig. 3f. For the control device (device C), the normalized UMR remains zero as a function of increasing charge current applied along the $b$-axis. On the other hand, the normalized UMR scales linearly with charge current density in device A and device B, which distinctly suggests that the unconventional UMR in these devices is due to the presence of $\sigma_z$, which is only allowed when a charge current is applied along the low-symmetry axis, i.e., $a$-axis, in WTe$_2$.

Note that such a non-linear magnetoresistance can also be induced by a thermal gradient due to Joule heating in the form of $|\nabla T| \propto I^2$ through thermoelectric effects such as anomalous Nernst effect and spin Seebeck effect[49,50]. The electric field due to thermoelectric effects has the form of $\boldsymbol{E}_{\text{TE}} \propto \boldsymbol{m} \times \nabla T$, therefore, to observe change in $R_{xx}$ for out-of-plane magnetization, it requires the presence of an in-plane thermal gradient ($y$-direction). We will provide three pieces of evidence to show that the thermoelectric effects are not dominant in our system as compared to the unconventional UMR. First, in our devices, the WTe$_2$ layer is conducting most of the charge current, and the thermal gradient is mostly out-of-plane and thus the in-plane thermal gradient is expected to be small[50]. Second, if the UMR is due to the thermoelectric effects, one would also expect a similar magnetoresistance effect when the current is applied along the $b$-axis. However, no clear signature of UMR is observed when current is applied along the $b$-axis in the control device (Fig. 4e). Finally, if there is a small in-plane ($y$-direction) thermal gradient in our system, it is expected to point in opposite directions at the two opposite edges of the WTe$_2$ since the in-plane $\nabla T$ should be roughly normal to the edge of the conducting region and pointing outward, i.e., towards the voltage probes. This will result in a sign change of such magnetoresistance when voltage probes on the opposite sides of the flakes are used to measure the longitudinal resistance. We have measured the unconventional UMR using voltage probes on opposite sides in device A and the UMR is of the same sign and similar magnitude (see supplementary Note 4). This unequivocally rules out the thermoelectric effects related origin of UMR and strongly suggests that the origin of observed unconventional UMR must be related to the intrinsic spin-dependent electron conduction affected by the exchange coupling interactions.

To understand unconventional UMR, we theoretically model the magneto-transport in WTe$_2$/CGT bilayers. The UMR characterizes the asymmetry in the longitudinal resistance when the direction of the applied electric field (or magnetization) is reversed. In the following, an electric field is assumed to be applied in the $x$-direction, but the results can be straightforwardly generalized to arbitrary directions. With the longitudinal resistivity defined as $\rho_{xx}(E_x) = E_x/j_x(E_x)$, the UMR ratio can be written as:

$$\text{UMR} = \frac{\rho_{xx}(E_x) - \rho_{xx}(-E_x)}{\rho_{xx}(E_x) + \rho_{xx}(-E_x)} = -\frac{j_x(E_x) + j_x(-E_x)}{j_x(E_x) - j_x(-E_x)} \quad (1),$$

If there is no asymmetry in the charge current for $\pm E_x$ field, i.e., $j_x(-E_x) = -j_x(E_x)$, then the UMR must vanish. This indicates that the UMR inherently belongs to the nonlinear response in the longitudinal direction. To the second order in $E_x$, the longitudinal current can be expanded as:

$$j_x(E_x) = j_x^{(1)} + j_x^{(2)} = \sigma_{xx}^{(1)} E_x + \sigma_{xx}^{(2)} (E_x)^2 + \cdots \quad (2)$$

The second-order conductivity $\sigma_{xx}^{(2)}$ introduces a nonlinear charge current $j_x^{(2)} \sim E_x^2$, which remains invariant after reversing the electric field, leaving an asymmetric longitudinal response $j_x(E_x) \neq j_x(-E_x)$ interpreted as the UMR. . For the case of $j_x^{(1)} \gg j_x^{(2)}$, the UMR defined in Eq.

(1) approximately equals to the normalized UMR, i.e., $\Delta R_{UMR}/R_0$, shown in the experimental results. After inserting Eq. (2) into Eq. (1), we can further simplify the UMR, arriving at a simple relation:

$$\text{UMR} \approx -\frac{\sigma_{xx}^{(2)}}{\sigma_{xx}^{(1)}} E_x \qquad (3)$$

The microscopic origin of the out-of-plane oriented spins in WTe$_2$ can be attributed to the layered spin Edelstein effect, which is consistent with the screw-axis and glide-plane symmetries in WTe$_2$ [16,36] (see Supplementary note 10). As the CGT is an insulating ferromagnet, considering only the first-adjacent WTe$_2$ layer is enough for a qualitatively modeling of the electron transport in CGT/WTe$_2$ bilayers[6]. Additionally, one would expect to observe a negligible UMR in the bulk WTe$_2$ sample which is consistent with our experimental data (see Supplementary note 5). We include an exchange coupling ($\Delta_{ex} \propto m_z$) and an interfacial Rashba SOC ($\alpha_R$) into the 1T$_d$-monolayer WTe$_2$ Hamiltonian[37] (see Methods and Supplementary note 8). With this minimal model, we first calculate the first-order conductivity for CGT/WTe$_2$ bilayer and the results are shown in Fig. 5a, where $\sigma_{xx}^{(1)}$ and $\sigma_{yy}^{(1)}$ only differ slightly and are even in the exchange coupling $\Delta_{ex}$, and the $x$ ($y$) axis has been chosen to be along the $a$ ($b$) axis of the WTe$_2$ crystal. Comparatively, the transverse conductivity, $\sigma_{xy}^{(1)}$ and $\sigma_{yx}^{(1)}$ accounting for the AHE, are odd functions of $\Delta_{ex}$, which agrees qualitatively with our experimental observations. Note that for our measured devices, an applied charge current density ($J_{DC} \sim 10^{10}$ A/m$^2$) corresponds to an electric field, $E \sim 10^5$ V/m, which is of the same order of $E$ that we used in our calculations (Fig. 5c).

It is important to mention that the second-order longitudinal conductivity, which accounts for the UMR, vanishes if the electric field is applied along the $y$ direction ($b$-axis), i.e., $\sigma_{yy}^{(2)} = 0$ as shown in Fig. 5b. In contrast, $\sigma_{xx}^{(2)}$ is finite such that when the electric field has a finite projection along the $x$ direction (i.e., the $a$-axis), a non-vanishing UMR should appear, agreeing well with experimental observations. In Fig.5c and Fig. 5d, we plot the UMR, calculated from Eq. (3), as functions of $B_z$ and $E_x$, respectively. Figure 5c shows that the UMR flips sign when the perpendicular magnetic field reverses direction, consistent with our experimental results. This behavior can be understood by the sign change of $\Delta_{ex}$ (Fig. 5b) caused by the magnetization switching of CGT. Note that for magnetic fields smaller than 50 mT, where CGT has low remanent magnetization due to either a reduced magnetic anisotropy or the formation of multi-domain, the results are not shown[32]. Because the magnetic field enters the WTe$_2$ Hamiltonian through the Zeeman coupling, which is usually much weaker than the exchange coupling, the UMR is almost independent of the applied magnetic field after the magnetization is saturated. We also found that the AHE (UMR) is suppressed (enhanced) with an increasing electron chemical potential around the edge of the valence band (see supplementary note 9), in qualitative agreement with our gate voltage dependent measurements of AHE and UMR in hole doped WTe$_2$/ CGT devices. This suggests that by tuning the electron chemical potential in WTe$_2$, the UMR can be maximized and explored in future devices with double gates and thinner h-BN dielectric for higher gate tunability. The UMR can also be phenomenologically understood by an intuitive symmetry consideration. From our calculation of nonequilibrium spin accumulation in the linear response to an applied electric field, we find that the $z$-component of the nonequilibrium spin accumulation $\sigma_z$ in WTe$_2$ is finite only when the electric field is applied along the $x$ direction, i.e., $a$-axis, (see supplementary note 10). Consequently, with a fixed magnetization along the $z$-direction, switching the current in the $x$-direction results in opposite $\sigma_z$ in WTe$_2$ which is subject to opposite exchange field from the CGT layer, thus introducing an asymmetric spin-dependent transport in WTe$_2$ that manifests as UMR. Previously, a nonreciprocal magnetoelectric effect has been observed in a monolayer WTe$_2$ coupled to a layered antiferromagnet, which originated from a quantum spin Hall effect in WTe$_2$ in the presence of magnetic proximity effect[51]. However, unconventional UMR in our devices with few-layer WTe$_2$ is not associated with helical edge states.

One of the outstanding challenges in the spintronics research is to realize a planar 2-terminal SOT-magnetic memory device based on bilayer systems[11,14], which requires that the out-of-plane magnetic state be electrically readable through 2-point measurements. So far, we have shown the utilization of the UMR to electrically read the magnetic state of CGT employing 4-point longitudinal measurements. Next, we show that the UMR can be employed to electrically read the magnetic state of CGT using a 2-point measurement configuration. To this end, we employ device A and measure the UMR response using a 2-point configuration (top panel in Fig. 5e) and the obtained $R_{xx}$ as a function of $B_z$ with positive and negative current is shown in Fig. 5e (middle panel). The extracted $R_{xx}^{asym}$ (lower panel in Fig. 5e) clearly shows that the perpendicular magnetization of CGT can be electrically detected by the unconventional UMR in 2-point measurements. Fig. 5f shows the current dependence of $\Delta R_{UMR}$ measured in 2-point configuration in device A, which is in good agreement with the 4-point measurement results. Note that the observed UMR is slightly larger in the 2-point method due to the increased length of the measurement channel, i.e., the spatial separation of voltage probes.

We have experimentally reported the discovery of an unconventional UMR in bilayer WTe$_2$/CGT heterostructures, which allows to read the up and down states of the PMA magnetic thin film through longitudinal resistance measurements. Our comprehensive experiments and theoretical modeling confirm that the observed UMR in bilayer WTe$_2$/CGT system originates from the interplay between the interfacial exchange coupling and the interfacial Rashba spin-orbit coupling, in the presence of the low-symmetry crystal structure of WTe$_2$. Our findings open an exciting opportunity to realize SOT-based two-terminal magnetic memory devices for the development of next-generation non-volatile and low-power consumption data storage technologies. Future studies to optimize and obtain a large unconventional UMR signal at room temperature will be crucial for practical device applications.

**Acknowledgments**
S.S. acknowledges the financial support from National Science Foundation (NSF) through Award No. ECCS-2208057, U.S. Office of Naval Research under Award No. N00014-23-1-2751, and the Center for Emergent Materials at The Ohio State University, a National Science Foundation (NSF) MRSEC, through Award No. DMR-2011876. J.K. acknowledges the financial support from U.S. Office of Naval Research under Award No. N00014-23-1-2751, the Center for Emergent Materials at The Ohio State University, an NSF MRSEC, through Award No. DMR-2011876 and the U.S. Department Office of Science, Office of Basic Sciences, of the U.S. Department of Energy under Award No. DE-SC0020323. J.T. and R.C. is supported by the Air Force Office of Scientific Research under grant FA9550-19-1-0307. J.Q.Y. acknowledges support from the U.S. Department of Energy, Office of Science, Basic Energy Sciences, Materials Sciences and Engineering Division. D.G.M. acknowledges support from the Gordon and Betty Moore Foundation's EPiQS Initiative through Grant No. GBMF9069. J. H. acknowledges the financial support from the Center of Emergent Materials, an NSF MRSEC, through Award No. DMR-2011876. Electron microscopy was performed at the Center for Electron Microscopy and Analysis at The Ohio State University. J.H.E. acknowledges the support for hBN crystal growth from the U. S. Office of Naval Research under award number N00014-22-1-2582. K.W. and T.T. acknowledge support from the JSPS KAKENHI (Grant Numbers 21H05233 and 23H02052) and World Premier International Research Center Initiative (WPI), MEXT, Japan. We also thank Ryan Muzzio for his help to prepare schematic of $WTe_2$ crystal structure shown in Fig. 1.

**Author Contributions**
S.S. and J.K. supervised the research. I.K. prepared the devices and performed the experiments with assistance from S.Y. J.T. and R.C. performed the theoretical calculations based on a tight binding model. G.C.O., M.Z., and J.H. performed the STEM measurements. R.R. carried out polarized Raman measurements. Y.J. and D.G.M. provided the bulk crystals of $WTe_2$. J.L., J. H. E., K.W., and T.T. provided the bulk h-BN crystals. All authors contributed to write the manuscript.

**Competing interests**
The authors declare no competing interests.

## Figure Captions

**Fig. 1. WTe$_2$/CGT devices and their characterization. a,** A model showing the crystal structure of WTe$_2$ with *a*-axis and *b*-axis labeled. The crystal is invariant (noninvariant) under a *bc* (*ac*) mirror operation. The mirror (M) and glide (G) planes are indicated by yellow dashed lines. **b,** Left: A schematic showing the generation of spin current with an out-of-plane spin polarization ($\sigma_z$), when a charge current (*J*) is applied along the *a*-axis of WTe$_2$. Right: A schematic depicting the concept of unconventional UMR in bilayer systems, i.e., change of longitudinal resistance ($R_{UMR}$) depending on the relative orientation of out-of-plane magnetization ($m_z$) and spin-current polarization ($\sigma_z$). When the magnetization is parallel (antiparallel) to the spin polarization, $R_{UMR}$ is negative (positive), leading to low (high) resistance state. **c,** Top: A schematic showing the side view of typical device. Bottom: An optical image of device A with the *a*-axis of WTe$_2$ aligned along the current electrodes, with CGT, WTe$_2$, graphite, and h-BN flakes outlined and labeled. The Pt electrodes are outlined with dashed black lines. Inset: An optical image of the WTe$_2$ flake used in device A. **d,** Angle-dependent polarized Raman spectral intensity at 163 cm$^{-1}$ of WTe$_2$ flake used in device A. **e,** Cross-sectional STEM image of the WTe$_2$/ CGT device B as viewed along the crystallographic *b*-axis of WTe$_2$. **f,** AFM line scan at the edges of the flakes, reflecting the thickness of WTe$_2$ and CGT. Inset: The full topography map obtained by AFM. **g,** 2-point longitudinal resistance as a function of temperature measured in a single device (device D) having a WTe$_2$/CGT region (blue curve) and a CGT region (orange).

**Fig. 2. Anomalous Hall effect in WTe$_2$/CGT devices. a,** The 4-point longitudinal resistance ($R_{xx}$) as a function of gate voltage ($V_g$) measured in Device A. The hysteresis in $R_{xx}$ is likely due to the ferroelectric behavior of WTe$_2$. **b,** The AHE hysteresis loops ($R_{yx}$ vs $B_z$), measured at 10 K at various gate voltages ($V_g$) when a charge current of $|I_{DC}| = 100\ \mu A$ is applied along the *a*-axis in Device A. Inset: A schematic showing the configuration for the transverse resistance ($R_{yx}$) measurement. **c,** The measured AHE resistance ($\Delta R_{AHE}$) as a function of the gate voltage ($V_g$) at 10 K. **d,** The AHE resistance ($\Delta R_{AHE}$) measured at $V_g = 0$ as a function of sample temperature. **e,** Top panel: The AHE hysteresis loops measured in device B at $+100\ \mu A$ ($-100\ \mu A$) are shown in orange (blue) curve, depicting no significant difference upon reversing the charge current polarity. Bottom panel: The antisymmetric part ($R_{yx}^{asym}$) and the symmetric part ($R_{yx}^{symm}$) of the transverse resistance is plotted in yellow and purple curve respectively. **f,** The obtained $\Delta R_{AHE}$ as a function of the gate voltage ($V_g$) when a charge current of $|I_{DC}| = 100\ \mu A$ is applied along the *a*-axis in Device B at 10 K. Inset: $R_{xx}$ as a function of $V_g$ in Device B, with similar hysteresis loop as observed in other device.

**Fig. 3. Unconventional unidirectional magnetoresistance in WTe$_2$/ CGT devices. a,** The raw longitudinal resistance ($R_{xx}^{raw}$) signal measured as a function of out-of-plane magnetic field ($B_z$) when a positive charge current of $+100\ \mu A$ is applied along the *a*-axis in Device A at 10 K. Inset to **(c)** shows a schematic diagram depicting the configuration for $R_{xx}$ vs $B_z$ measurements. The orange (yellow) lines are fit to the saturated region at the positive (negative) B-field side to subtract the background not associated with UMR. **b,** Top panel: The obtained $R_{xx}$, after subtracting a trivial background at saturated magnetization regions as shown in **(b)**, when a positive (orange curve) and a negative (blue curve) charge current. Bottom panel: The antisymmetric part ($R_{xx}^{asym}$) and the symmetric part ($R_{xx}^{symm}$) of the longitudinal resistance is obtained and is plotted

in yellow and purple curve respectively. **c,** $R_{xx}$ vs $B_z$ hysteresis loops, associated with unconventional UMR, are measured at 10 K at various gate voltages ($V_g$). **d,** The magnitude of UMR ($\Delta R_{UMR}$), as defined in lower panel of **(b)**, is plotted as a function of $V_g$ at 10 K. **e,** $\Delta R_{UMR}$ as a function of the temperature with a charge current of $|I_{DC}| = 100\ \mu A$ and $V_g = 0\ V$. **f,** The magnitude of measured UMR at positive (orange curve) and negative (blue curve) charge current, i.e., $\Delta R_{UMR}^{+/-}$, as a function of the charge current magnitude ($|I_{DC}|$) in Device A at 10 K. **g,** The measured $R_{xx}^{raw}$ as function of $B_z$ when a negative charge current of $-300\ \mu A$ is applied along the $a$-axis in Device B. The orange (yellow) curve is fit to the background at saturated region at the positive (negative) field side. **h,** Top panel: The obtained $R_{xx}$, after subtracting a trivial background at saturated magnetization regions as shown in **(g)**, at positive (orange curve) and a negative (blue curve) charge current. Bottom panel: The $R_{xx}^{asym}$ and $R_{xx}^{symm}$ as a function of $B_z$, obtained from the top panel, are plotted in yellow and purple curve respectively. **i,** The obtained $\Delta R_{UMR}$ as a function of $V_g$ at 10 K when a charge current of $|I_{DC}| = 300\ \mu A$ is applied.

**Fig. 4. Absence of unconventional unidirectional magnetoresistance in control device. a,** Top panel: An optical image of the WTe$_2$ flake used in device C, with $a$-axis and $b$-axis orientation labeled. Bottom panel: An optical image of device C with the $b$-axis of WTe$_2$ aligned to the current electrodes, with flake of CGT, WTe$_2$, graphite, and h-BN outlined and labeled. The Pt electrodes are outlined with dashed black lines. **b,** Angle-dependent polarized Raman spectral intensity at 163 cm$^{-1}$ of WTe$_2$ flake, shown in **(a)**, used in device C, confirming that the $b$-axis of the WTe$_2$ flake is aligned parallel to the current electrodes. **c,** The measured $R_{xx}$ as a function of $V_g$ in Device C at 10K showing a hysteresis, which is likely due to the ferroelectric switching in WTe$_2$. **d,** The AHE hysteresis loops ($R_{yx}$ vs $B_z$), measured at 10 K at $V_g$ when a charge current of $|I_{DC}| = 200\ \mu A$ is applied along the $b$-axis in Device C and magnetic field applied in the out-of-plane direction. Inset: A schematic showing the configuration for the $R_{yx}$ vs $B_z$ measurements. **e,** The obtained asymmetric longitudinal resistance ($R_{xx}^{asym}$) hysteresis loops at various $V_g$ when a charge current of $|I_{DC}| = 200\ \mu A$ is applied along the $b$-axis in device C at 10 K shows no signature of the presence of unconventional UMR. Inset: A schematic depicting the configuration for $R_{xx}$ vs $B_z$ measurements. **f,** The obtained normalized UMR magnitude ($|\Delta R_{UMR}/R_0|$), where $R_0$ is the magnetic field independent part of the longitudinal resistance, as a function of the charge current density magnitude $|J_{DC}|$ in device A, B, and C. A linear current density dependence of normalized UMR is only observed in device A and B, wherein the current is applied along the $a$-axis of WTe$_2$.

**Fig. 5. Theoretical modeling of unidirectional magnetoresistance and anomalous Hall effect. a, b,** First-order (**a**) and second order (**b**) conductivity as a function of exchange coupling. The transverse conductivity $\sigma_{xy}^{(1)} = -\sigma_{yx}^{(1)}$ accounts for the AHE and UMR is proportional to the second-order conductivity, i.e., $\sigma_{xx}^{(2)}$ and $\sigma_{yy}^{(2)}$. **c,d,** UMR as a function of out-of-plane magnetic field, $B_z$ (**c**) and electric field, $E$ (**d**) when the electric field is applied along the $a$-axis (red curve) and along the $b$-axis (blue curve) of WTe$_2$.. **e,** Top panel: A schematic showing the measurement configuration to read the out-of-plane magnetic state of CGT employing 2-point longitudinal resistance ($R_{xx}$). Middle panel: The measured 2-point $R_{xx}$ when a positive (orange curve) and a negative (blue curve) charge current is applied along the $a$-axis of WTe$_2$ in Device A. The step-like change in 2-point $R_{xx}$ is due to UMR switching sign when the current is reversed. Bottom panel: The antisymmetric part ($R_{xx}^{asym}$) and the symmetric part ($R_{xx}^{symm}$) of the 2-point $R_{xx}$ is plotted in yellow and purple curve respectively, which clearly shows that the out-of-plane magnetization of CGT can be electrically read through UMR. **f,** The measured magnitude of UMR ($\Delta R_{UMR}^{+/-}$), obtained from 2-point $R_{xx}$ measurements, as a function of the charge current magnitude $|I_{DC}|$ when a positive (orange curve) a negative (blue curve) charge current is applied along the $a$-axis of WTe$_2$ in Device A at 10 K.

## Methods

**Device Fabrication.** WTe$_2$ and hexagonal boron nitride (h-BN) crystals were prepared by previously published procedures[31,52]. CGT single crystals were purchased from HQ Graphene. Mechanical exfoliation of WTe$_2$, h-BN, graphite, and CGT was performed on separate silicon wafers with 300 nm of SiO$_2$ inside a glovebox filled with Ar gas. Flakes were selected through optical investigation through a microscope. WTe$_2$ flakes that have well-defined and straight edges were used because the *a*-axis tends to be along straight edges[39]. For most of the devices, before the heterostructure fabrication, the Pt electrodes were defined on a separate Si/ SiO$_2$ substrate using electron beam lithography (EBL) and sputtering deposition with a polymethyl methacrylate (PMMA) resist. The electrodes contacting the heterostructure were composed of Pt (6 nm) or Ti (1-3 nm)/ Pt (6-7 nm). The Pt electrodes are then connected by Cr (5 nm)/ Au (110 nm) electrodes for wire bonding pads, prepared using electron beam lithography (EBL) and electron beam deposition with a PMMA/ methylmethacrylate (MMA) bilayer resist.

The heterostructure was fabricated using a custom transfer tool inside a glovebox filled with Ar gas. A transfer slide consisting of a polydimethylsiloxane (PDMS) slab and a thin film of polycarbonate (PC) was used for picking up hBN, CGT, and WTe$_2$ in that order and then putting the stack on the Pt electrodes. For devices with a top gate, an additional graphite flake is picked up in the beginning to contact the pre-patterned Pt electrodes. As for devices sandwiched between two h-BN flakes, a graphite/ h-BN is first transferred to a Si/ SiO$_2$ substrate. Pt electrodes composed of Ti (1 nm)/ Pt (6 nm) were patterned on top of the h-BN using electron beam lithography (EBL) and electron beam deposition with a PMMA/ MMA bilayer resist. The Pt electrodes were connected to the Au electrodes by the same method mentioned previously. Finally, a stack consisting of WTe$_2$/CGT/h-BN was transferred on top of the Pt electrodes to complete a device. For all UMR devices, the Pt electrodes and the substrate were cleaned by atomic force microscopy in contact mode using $\mu$masch HQ:NSC15/Al BS tips and a gentle oxygen plasma before the heterostructure was transferred to the electrodes to ensure the interface quality.

**Electrical measurements.** The electrical measurements were performed at variable temperatures in high vacuum ($< 10^{-5}$ mtorr) conditions. An electromagnet was rotated such that the magnetic field could be applied in both in- and out-of-plane directions of the device. A Keithley 6221 current source and two Keithley 2182A nanovoltmeters are used for simultaneous DC measurements for both transverse and longitudinal resistances. A Keithley 2400 source meter is used for applying gate voltage to the device. For each magnetic field hysteresis loop, the transverse and longitudinal (4-point or 2-point) voltages are measured at each magnetic field when a constant DC charge current is applied. The antisymmetric and symmetric part of the resistance is obtained by measuring the hysteresis loops at positive and negative currents separately. We calculate the antisymmetric and symmetric part of the resistance as follows $R_{xx(yx)}^{symm(asym)} = \frac{R_{xx(yx)}(+I_{DC}) \pm R_{xx(yx)}(-I_{DC})}{2}$. The voltage response that is linear (quadratic) to the current is an odd (even) function of the current, while the resistance ($R \equiv V/I$) of such responses will be an even (odd) function of the current. The antisymmetric (symmetric) part of the resistance will eliminate the odd (even) terms and therefore the linear (quadratic) current-voltage responses. We utilize this method to better quantify the observed unconventional UMR. Symmetric parts were taken for the measurements of AHE, since that will also eliminate thermal effects that are usually quadratic current responses.

**Theory calculations.** The interface of CGT/WTe$_2$ bilayer heterostructure is minimally modeled by introducing the exchange coupling and interfacial Rashba SOC into the 1T$_d$-monolayer WTe$_2$ Hamiltonian[37]:

$$H(k_x, k_y) = \epsilon_0(k_x, k_y) + \beta \sin(k_y a_y)\sigma_0 \otimes \tau_0 + \eta \sigma_0 \otimes \tau_x + H_{soc} + H_{ex} + H_R,$$

where $\boldsymbol{\sigma}$ and $\boldsymbol{\tau}$ the vectors of Pauli matrices for the spin and sublattice degrees of freedom, respectively. Here, $\epsilon_0(k_x, k_y) = m_p[4 - 2\cos(k_x a_x) - 2\cos(k_y a_y)]\sigma_0 \otimes \tau_0 + m_d[4 - 2\cos(k_x a_x) - 2\cos(k_y a_y) + \delta]\sigma_0 \otimes \tau_z$, $H_{ex} = -\Delta_{ex}\sigma_z \otimes \tau_0$, $H_R = -\alpha_R[\sin(k_y a_y)\sigma_x \otimes \tau_0 - \sin(k_x a_x)\sigma_y \otimes \tau_0]$ and $H_{soc} = \Lambda_x \sin(k_y a_y)\sigma_x \otimes \tau_x + \Lambda_y \sin(k_x a_x)\sigma_y \otimes \tau_x + \Lambda_z \sin(k_x a_x)\sigma_z \otimes \tau_x$. The physical meaning of each term and the specific values of the parameters are detailed in the supplementary information.

In the metallic regime, the extrinsic (or Fermi-surface) contribution for the first- and second-order conductivity can be calculated by:

$$\sigma_{xx}^{(1)} = e^2\hbar \sum_n \int_{BZ} \frac{d^2\boldsymbol{k}}{(2\pi)^2} (\hat{v}_x^{nn})^2 \frac{\Gamma^2}{\pi[(\epsilon_n - \mu_F)^2 + \Gamma^2]^2},$$

$$\sigma_{xx}^{(2)} = -e^3 \sum_n \int_{BZ} \frac{d^2\boldsymbol{k}}{(2\pi)^2} \hat{v}_x^{nn}(\partial_{k_x}\hat{v}_x)^{nn} \frac{\Gamma^2}{\pi[(\epsilon_n - \mu_F)^2 + \Gamma^2]^2},$$

where $\Gamma$ represents the disorder broadening, $\hbar$ is the reduced Planck constant, $\hat{v}_x = \partial_{k_x}\hat{H}/\hbar$ is the velocity operator with matrix element being $\hat{v}_x^{nn} = \langle n|\hat{v}_x|n\rangle$, and the summation runs over all the energy bands. Since the Lorentzian function $\Gamma^2/\pi[(\epsilon_n - \mu)^2 + \Gamma^2]^2$ with a broadening $\Gamma$ is sharply centered around the chemical potential $\mu_F$, only the bands near the Fermi surface contribute to the longitudinal conductivity.

On the other hand, the anomalous Hall current $j_x^{(1)} = \sigma_{xy}^{(1)} E_y$ can be characterized by the first-order transverse conductivity $\sigma_{xy}^{(1)}$, which can be calculated from the intrinsic Fermi-sea contribution:

$$\sigma_{xy}^{(1)} = -2e^2\hbar \sum_n f(\epsilon_n, \mu_F) \sum_{m \neq n} \int_{BZ} \frac{d^2\boldsymbol{k}}{(2\pi)^2} \text{Im}[\hat{v}_y^{nm}\hat{v}_x^{mn}] \frac{(\epsilon_n - \epsilon_m)^2 - \Gamma^2}{[(\epsilon_n - \epsilon_m)^2 + \Gamma^2]^2}$$

Here, $f(\epsilon_n, \mu_F)$ is the Fermi-Dirac distribution function. Therefore, all the occupied bands below the Fermi surface contribute to the anomalous Hall conductivity. Note that the skew scattering and side jump mechanisms, which both belong to the extrinsic contribution, are not included in the above formula with the approximation of constant relaxation time $\tau = \hbar/2\Gamma$.

Similarly, the nonequilibrium spin density $\delta S_\mu = \chi_{\mu\nu} E_\nu$ induced by an applied electric field can be calculated from the extrinsic (*i.e.*, intraband) contribution given by the Kubo formula:

$$\delta S_\mu = e\hbar \sum_n \int_{BZ} \frac{d^2\boldsymbol{k}}{(2\pi)^2} \hat{S}_\mu^{nn}\hat{v}_\nu^{nn} \frac{\Gamma^2}{\pi[(\epsilon_n - \mu_F)^2 + \Gamma^2]^2},$$

where $\hat{S} = \hbar\boldsymbol{\sigma}/2$ is the spin operator. Note that we have ignored the intrinsic (*i.e.*, interband) contribution, which is overwhelmed by the extrinsic contribution in the metallic regime.

**Data availability**
All the data supporting the findings of this study are available in the main text and its Supplementary Information. Further information is available from the corresponding author on reasonable request.

**Figure 1**

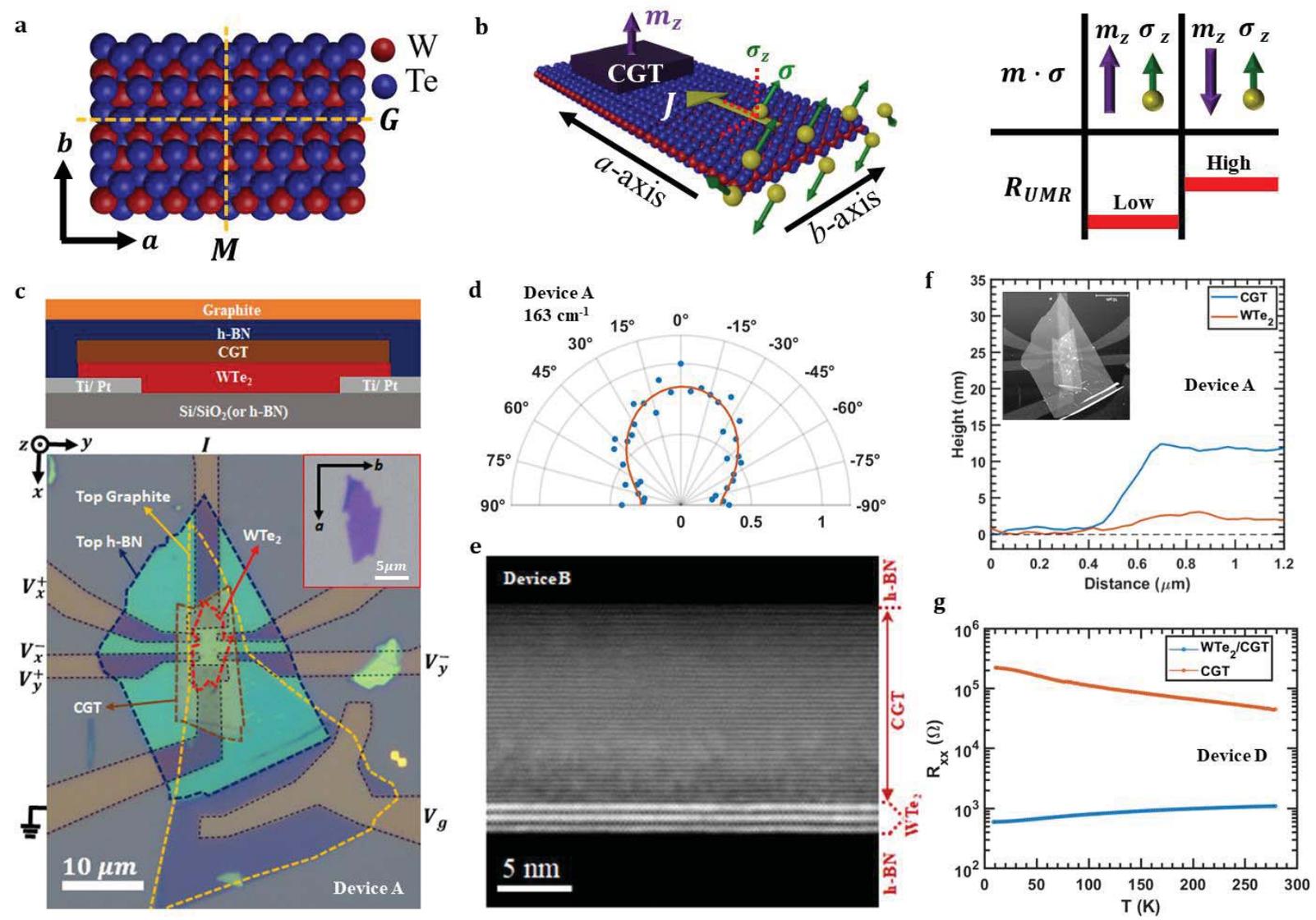

**Figure 2**

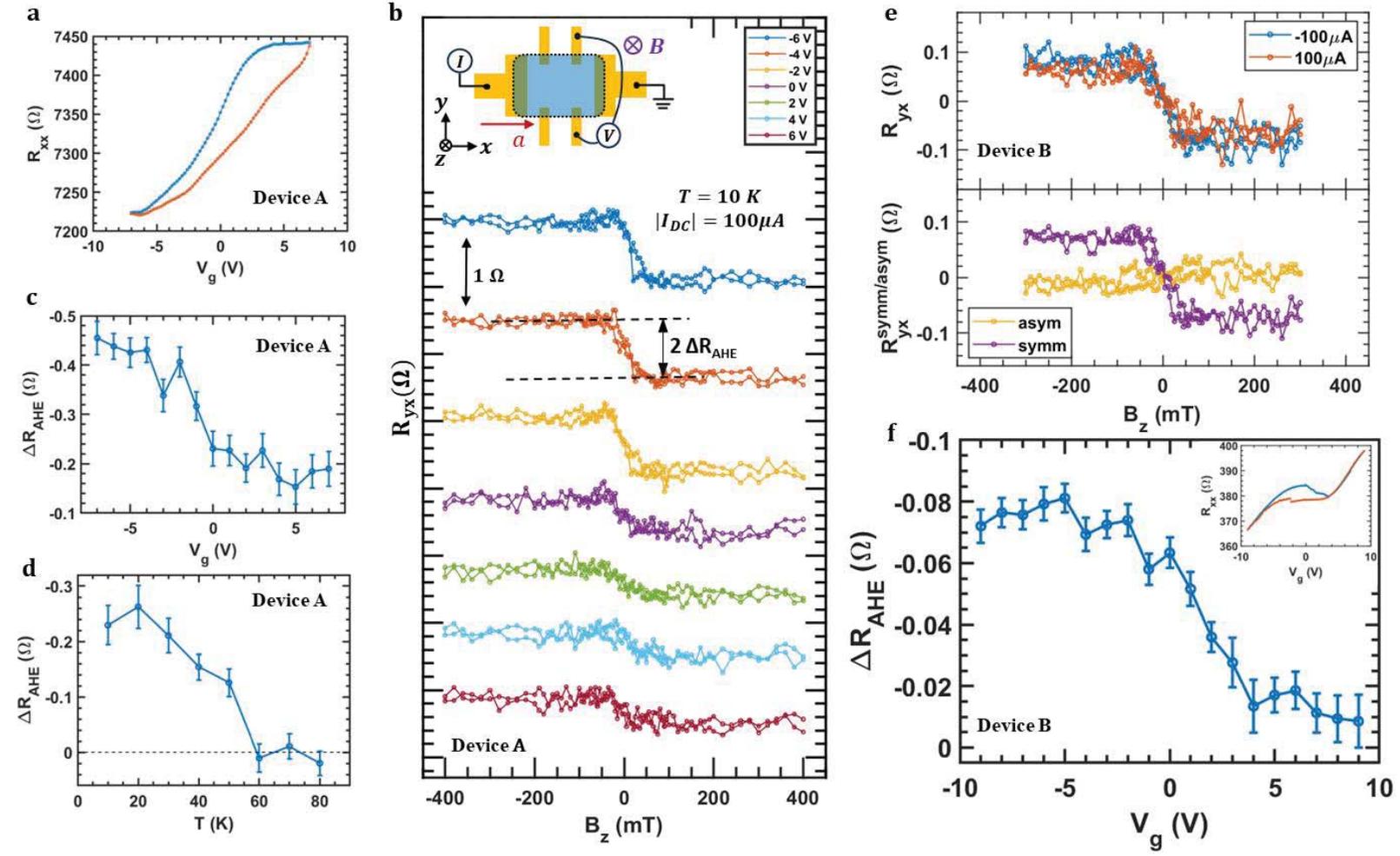

**Figure 3**

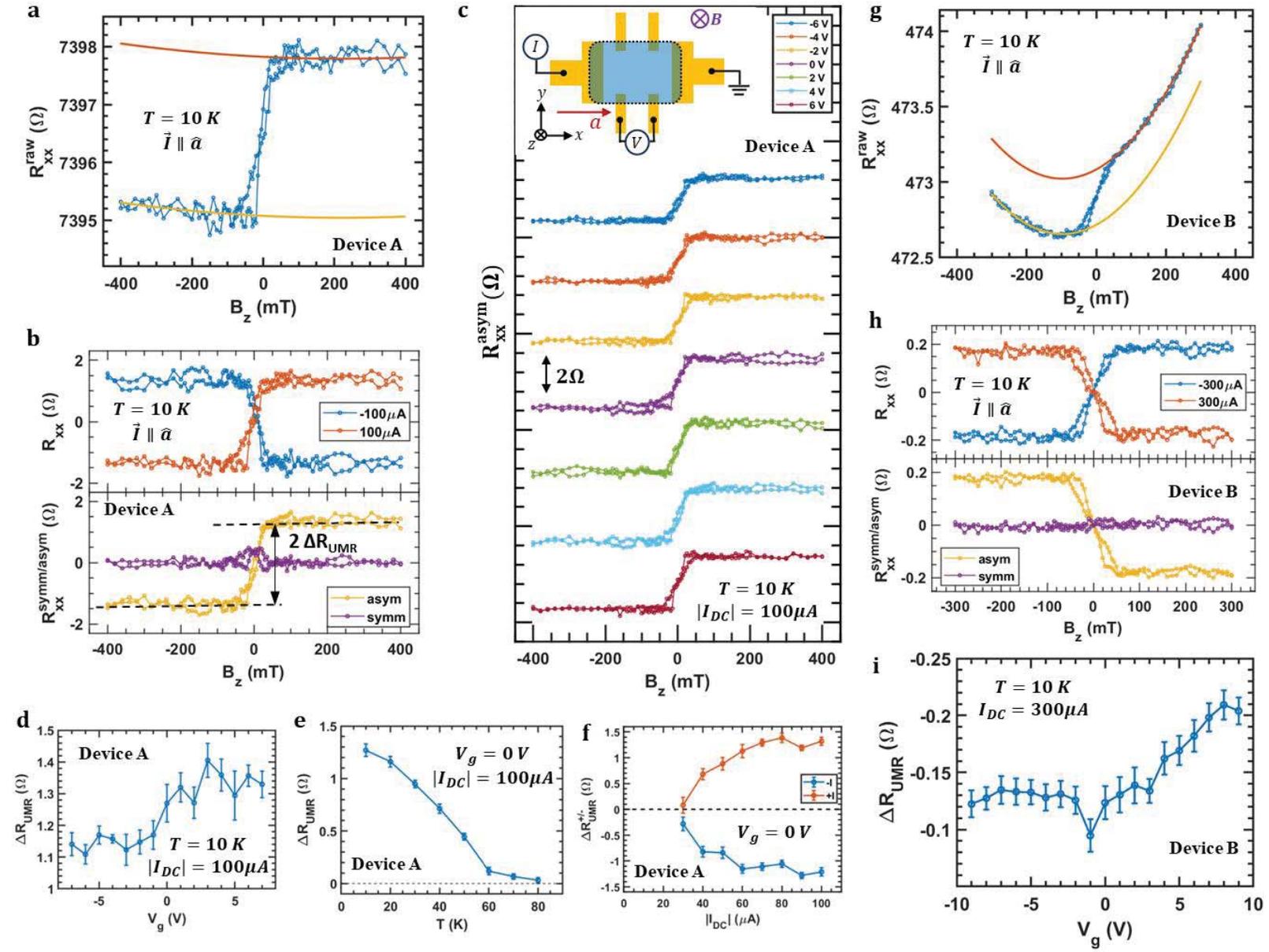

**Figure 4**

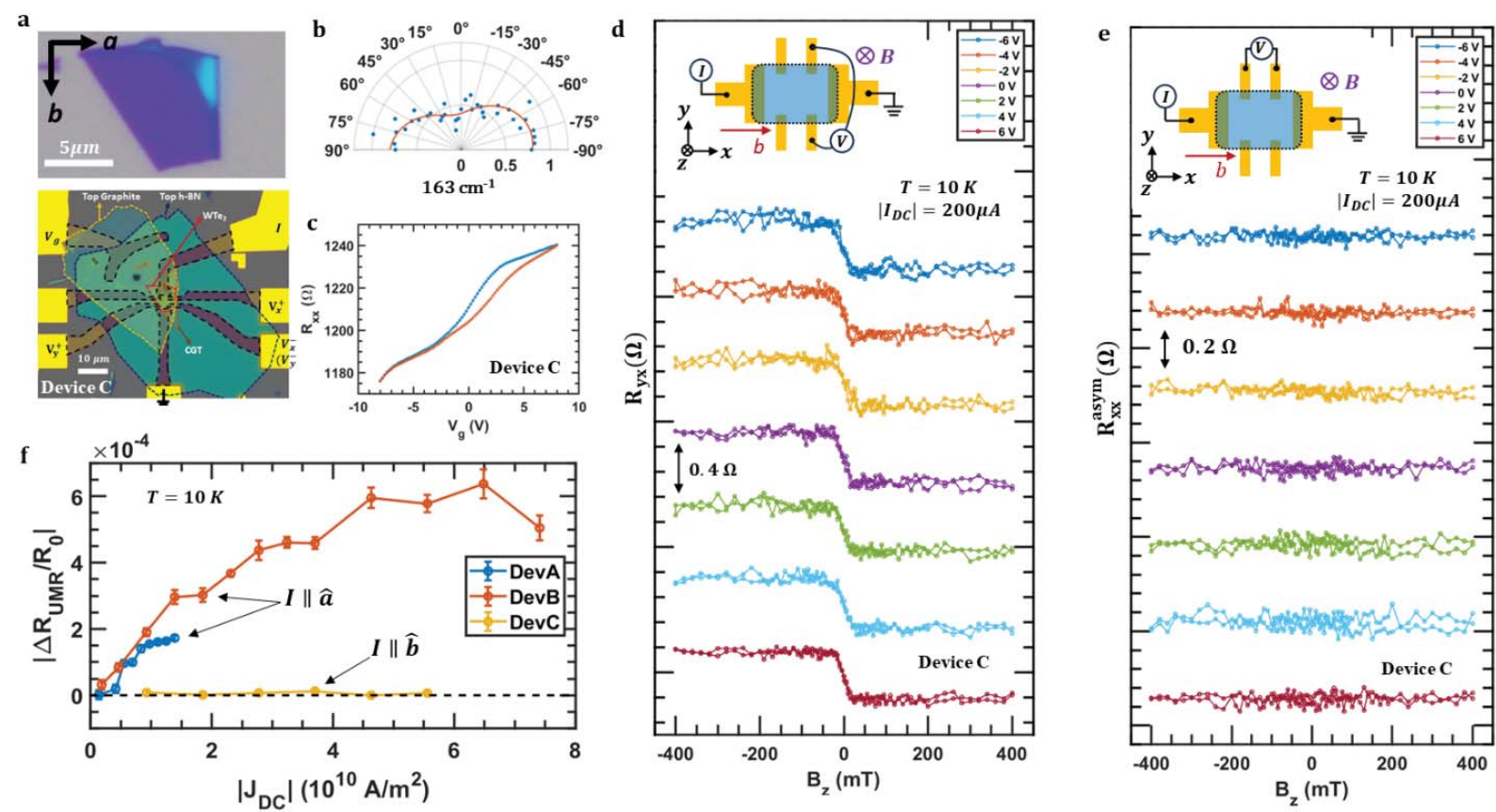

**Figure 5**

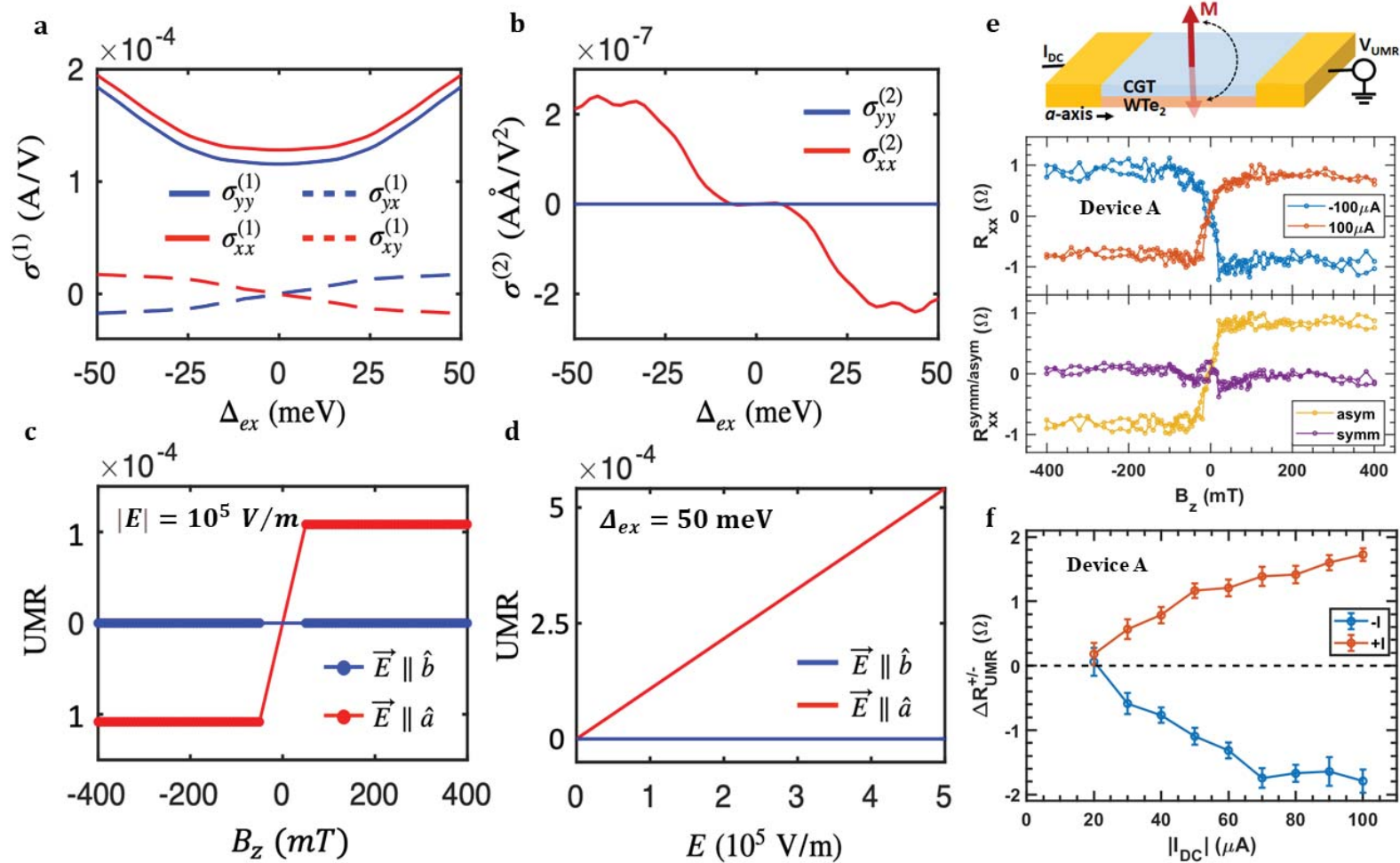

Supplementary Information for

# Unconventional Unidirectional Magnetoresistance in vdW Heterostructures

I-Hsuan Kao[1], Junyu Tang[2], Gabriel Calderon Ortiz[3], Menglin Zhu[3],
Sean Yuan[1], Rahul Rao[4], Jiahan Li[5], James H. Edgar[5], Jiaqiang Yan[6,7],
David G. Mandrus[6,7], Kenji Watanabe[8], Takashi Taniguchi[9],
Jinwoo Hwang[3], Ran Cheng[2,10,11], Jyoti Katoch[1], and Simranjeet Singh[1,*]

[1]Department of Physics, Carnegie Mellon University, Pittsburgh, PA, 15213, USA

[2]Department of Physics and Astronomy, University of California,
Riverside, California 92521, USA

[3]Department of Materials Science and Engineering, The Ohio State University,
Columbus, OH 43210, USA

[4]Materials and Manufacturing Directorate, Air Force Research Laboratory,
Wright-Patterson Air Force Base, Dayton, OH, 45433, USA

[5]Tim Taylor Department of Chemical Engineering, Kansas State University,
Manhattan, Kansas 66506, USA

[6]Materials Science and Technology Division, Oak Ridge National Laboratory,
Oak Ridge, Tennessee 37831, USA

[7]Department of Materials Science and Engineering, The University of Tennessee,
Knoxville, TN 37996, USA

[8]Research Center for Electronic and Optical Materials,
National Institute for Materials Science, 1-1 Namiki, Tsukuba 305-0044, Japan

[9]Research Center for Materials Nanoarchitectonics,
National Institute for Materials Science, 1-1 Namiki, Tsukuba 305-0044, Japan

[10]Department of Electrical and Computer Engineering, University of California,
Riverside, California 92521, USA

[11]Department of Materials Science and Engineering, University of California,
Riverside, California 92521, USA

*Email: simranjs@andrew.cmu.edu

**Contents**

**Note 1.   Device Fabrication**

Mechanical exfoliation of WTe$_2$, hBN, graphite, and Cr$_2$Ge$_2$Te$_6$ (CGT) was performed on separate silicon wafers with 300 nm of SiO$_2$ inside of an Ar environment. Flakes were selected through optical investigation through a microscope. WTe$_2$ flakes that have well-defined and straight edges were used because the *a*-axis tends to be along them[1]. For most of the devices, before the heterostructure fabrication, the Pt electrodes were defined on a separate Si/ SiO$_2$ chip substrate using electron beam lithography (EBL) and sputtering deposition with a polymethyl methacrylate (PMMA) resist. The electrodes contacting the heterostructure were composed of Pt(6 nm) or Ti(1-3 nm)/ Pt(6-7 nm). The Pt electrodes are then connected by Cr(5 nm)/ Au (110 nm) electrodes for wire bonding pads, prepared using electron beam lithography (EBL) and electron beam deposition with a PMMA/ methylmethacrylate (MMA) bilayer resist. The heterostructure was fabricated using a custom transfer tool in an Ar environment. A transfer slide consisting of a polydimethylsiloxane (PDMS) slab and a thin film of polycarbonate (PC) was used for picking up hBN, CGT, and WTe$_2$ in that order and then putting the stack on the Pt electrodes. For devices with a top gate, an additional graphite flake is picked up in the beginning to contact the pre-patterned Pt electrodes.

As for devices sandwiched between two hBN flakes, a graphite/ hBN is first transferred to a Si/ SiO$_2$ substrate. Pt electrodes composed of Ti(1 nm)/ Pt(6 nm) were patterned on top of the hBN using electron beam lithography (EBL) and electron beam deposition with a PMMA/ MMA bilayer resist. The Pt electrodes were connected to the Au electrodes by the same method mentioned previously. Finally, a stack consisting of WTe$_2$/ CGT/ hBN was transferred on top of the Pt electrodes to complete a device. For all the Unidirectional magnetoresistance (UMR) devices, the Pt electrodes and the substrate were cleaned by atomic force microscopy (AFM) in contact mode using $\mu$masch HQ:NSC15/Al BS tip and a gentle oxygen plasma before the heterostructure was transferred to the electrodes to ensure the interface quality.

**Note 2. Thickness, Interface, and Crystallographic Orientation of WTe$_2$/CGT Devices**

Six WTe$_2$/CGT van der Waals (vdW) heterostructures were fabricated and named device A to F (Fig. S1-6). The WTe$_2$ alignment, electrode geometry, and thickness parameters of WTe$_2$, CGT, top hBN (thBN), and bottom hBN (bhBN) flakes in each WTe$_2$/CGT device are summarized in Table S1. The results presented in the main paper come from device A, B, C, and D. Most of the devices are made into Hall bar with *a*-axis of WTe$_2$ aligned the to current channel for accurate four-point electric measurements of UMR and Anomalous Hall effect (AHE). We observed unconventional UMR in device A and B, where thin WTe$_2$ (bilayer and trilayer) and thicker CGT ( 10 nm) were used. In device C, we intentionally aligned the *b*-axis of WTe$_2$ to the current channel and showed that the unconventional UMR disappeared due to the absence of out-of-plane spins. In Device D, we intentionally made a double Hall cross-electrode geometry so that we have one region with WTe$_2$/ CGT and one region with only CGT. This is for checking the longitudinal resistance between WTe$_2$/ CGT and CGT to ensure most of the current is flowing through the WTe$_2$ layer. In device D, charge current can be applied along both *a* and *b*-axis due to the Hall cross geometry, where we observed AHE in both cases. We did not measure UMR in device D. In Device E, we were not able to observe both AHE and UMR. We suspected that this is due to thin CGT (1.4 nm) used in this device, resulting in the reduction of exchange coupling interaction. In Device F, we observed AHE but not UMR, which might be due to thicker WTe$_2$ (6.3 nm). As the unconventional UMR originates from the first WTe$_2$ layer neighboring the CGT, the magnitude of such effect is expected to be reduced when the thickness of WTe$_2$ increases.

To determine the thickness, crystallographic orientation, and interface quality of the devices, we utilize AFM, polarized Raman spectroscopy, and scanning transmission electron microscopy (STEM). All the devices were examined by AFM in the tapping mode using $\mu$masch HQ:

NSC15/ Al BS tip to investigate the thickness of flakes used in each device. For device A and D, cross-sectional STEM imaging was used to investigate the interface quality and the thickness of CGT and WTe$_2$. The crystallographic orientation of WTe$_2$ in device A, B, C, and E were confirmed by angle-dependent Polarized Raman measurements. The angle between the laser polarization and the current channel is defined by $\phi_{Raman}$. Since the Raman shift peaks of WTe$_2$[1] at 83, 112, 117, 137, and 212 cm$^{-1}$ overlap with the Raman shift peaks of CGT[2,3] at 85, 111, 136, and 213 cm$^{-1}$, we focused on the Raman shift peak of WTe$_2$ at 163 cm$^{-1}$ to determine the crystallographic orientations of the WTe$_2$ flake. The Raman intensity at 163 cm$^{-1}$ is maximized when laser polarization is parallel to the *a*-axis of WTe$_2$. In figure S1b, 2b, 5b, the maxima at $\phi_{Raman} = 0$ confirmed that the *a*-axis of WTe$_2$ in device A, B, and E is parallel to the current channel. As for device C (figure S3), the minima at $\phi_{Raman} = 0$ shows that the *b*-axis of WTe$_2$ is parallel to the current channel.

**Table S1** | The WTe$_2$ alignment, electrode geometry, and thickness parameters of WTe$_2$, CGT, thBN, and bhBN flakes in WTe$_2$/ CGT devices.

| Device | $t_{WTe_2}$ (nm) | $t_{CGT}$ (nm) | $t_{thBN}$ (nm) | $t_{bhBN}$ (nm) | WTe$_2$ axis alignment | Electrode geometry |
|---|---|---|---|---|---|---|
| A | 1.6 | 11.0 | 23.9 | N/A | $a$-axis | Hall bar |
| B | 2.2 | 13.2 | 20.5 | 30.4 | $a$-axis | Hall bar |
| C | 2.3 | 8.1 | 18.0 | N/A | $b$-axis | Hall bar |
| D | 7.4 | 6.5 | 29.9 | N/A | $a$ & $b$-axis | Hall cross |
| E | 1.4 | 1.4 | 17.9 | 21.6 | $a$-axis | Hall bar |
| F | 6.3 | 10.9 | 35.2 | N/A | $a$-axis | Hall bar |

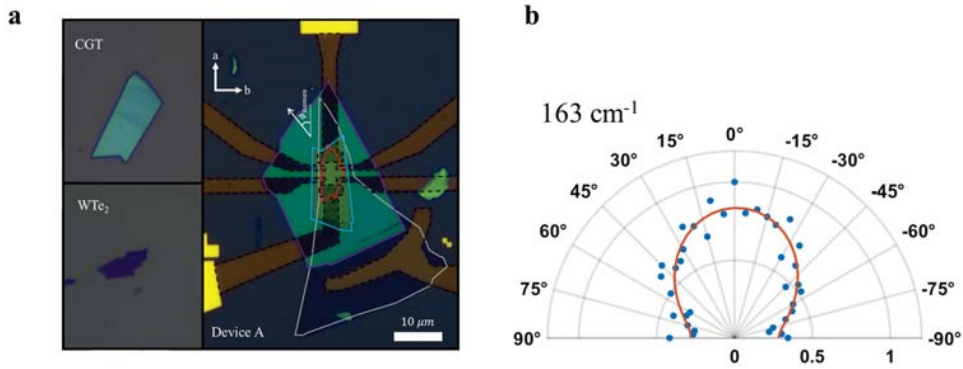

**Figure S1** | **The optical images and axis characterization of device A**. **a,** the optical images of CGT and WTe$_2$ flakes (left) and the complete stack (right), where the outlines of Pt (dashed black), WTe$_2$ (red), CGT (cyan), top hBN (purple), and top graphite (grey) are indicated. The axes of WTe$_2$ are labeled based on the polarized Raman results. **b,** Raman data was collected as a function of polarization angle where the $\phi_{Raman} = 0°$ is aligned to the current channel (as shown in **a**). The intensity of the Raman peak at 163 cm$^{-1}$ as a function of polarization angle is plotted as a polar plot with a $cos^2(\phi_{Raman})$ fit. The maxima around $0°$ confirms that the $a$-axis of WTe$_2$ is aligned along the current channel.

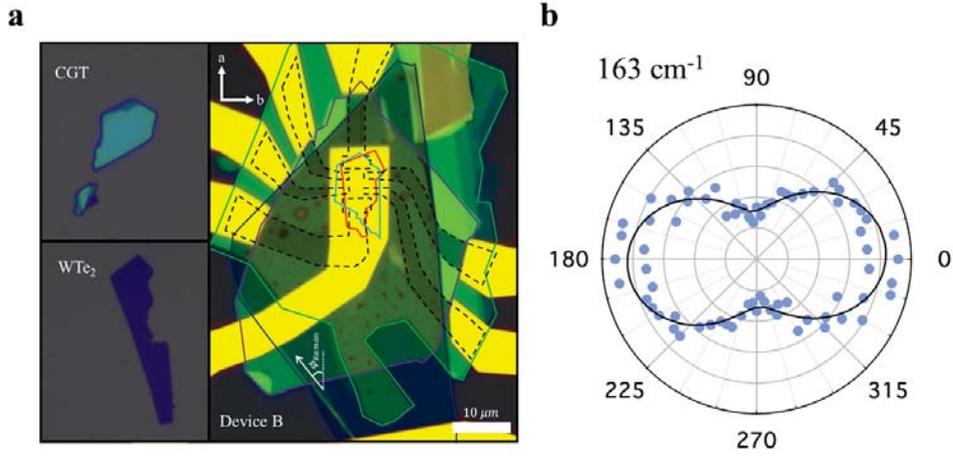

**Figure S2 | The optical images and axis characterization of device B**. **a,** the optical images of CGT and WTe$_2$ flakes (left) and the complete stack (right), where the outlines of bottom graphite (blue), bottom hBN (green), Pt (dashed black), WTe$_2$ (red), CGT (cyan), and top hBN (purple) are indicated. The axes of WTe$_2$ are labeled based on the polarized Raman results. **b,** Raman data was collected as a function of polarization angle where the $\phi_{Raman} = 0°$ is aligned to the current channel (as shown in **a**). The intensity of the Raman peak at 163 cm$^{-1}$ as a function of polarization angle is plotted as a polar plot with a $cos^2(\phi_{Raman})$ fit. The maxima around 0° confirms that the $a$-axis of WTe$_2$ is aligned along the current channel.

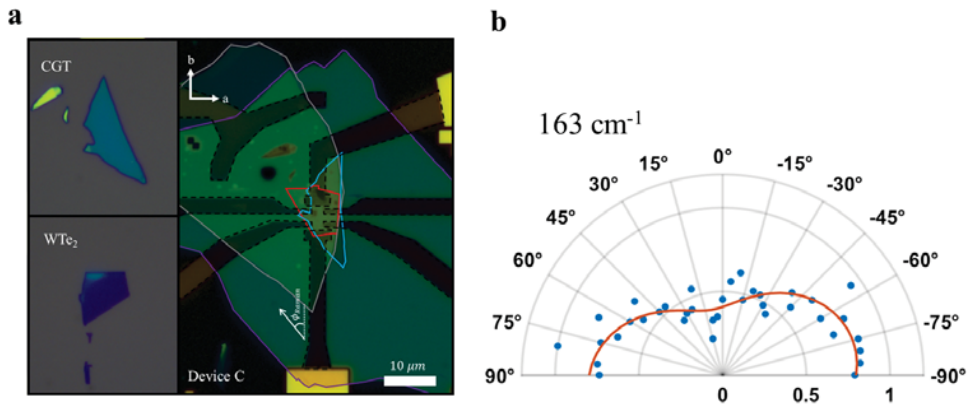

**Figure S3 | The optical images and axis characterization of device C**. **a,** the optical images of CGT and WTe$_2$ flakes (left) and the complete stack (right), where the outlines of Pt (dashed black), WTe$_2$ (red), CGT (cyan), top hBN (purple), and top graphite (grey) are indicated. The axes of WTe$_2$ are labeled based on the polarized Raman results. **b,** Raman data was collected as a function of polarization angle where the $\phi_{Raman} = 0°$ is aligned to the current channel (as shown in **a**). The intensity of the Raman peak at 163 cm$^{-1}$ as a function of polarization angle is plotted as a polar plot with a $cos^2(\phi_{Raman})$ fit. The minima around 0° confirms that the $b$-axis of WTe$_2$ is aligned along the current channel.

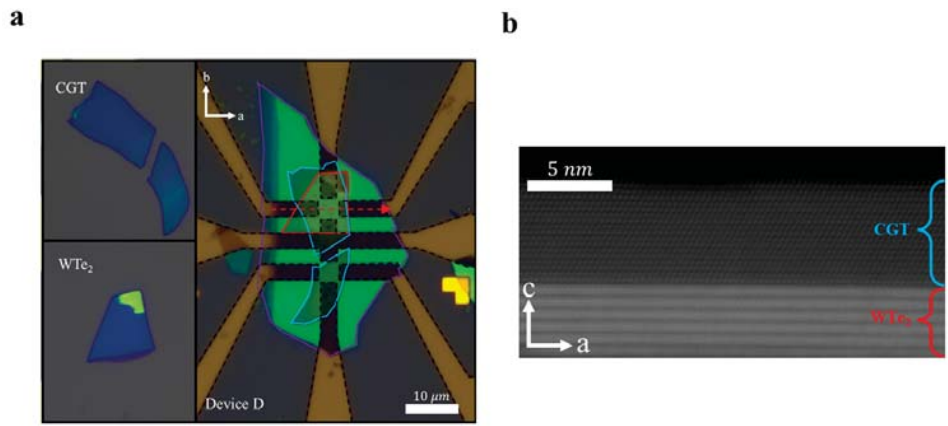

**Figure S4 | The optical images and axis characterization of device D**. **a,** the optical images of CGT and WTe$_2$ flakes (left) and the complete stack (right), where the outlines of Pt (dashed black), WTe$_2$ (red), CGT (cyan), and top hBN (purple) are indicated. The axes of WTe$_2$ are labeled based on the STEM imaging results. **b,** The cross-sectional STEM of device D taken along the red dashed arrow in **a**, showing that the cut is along the $a$-axis of WTe$_2$.

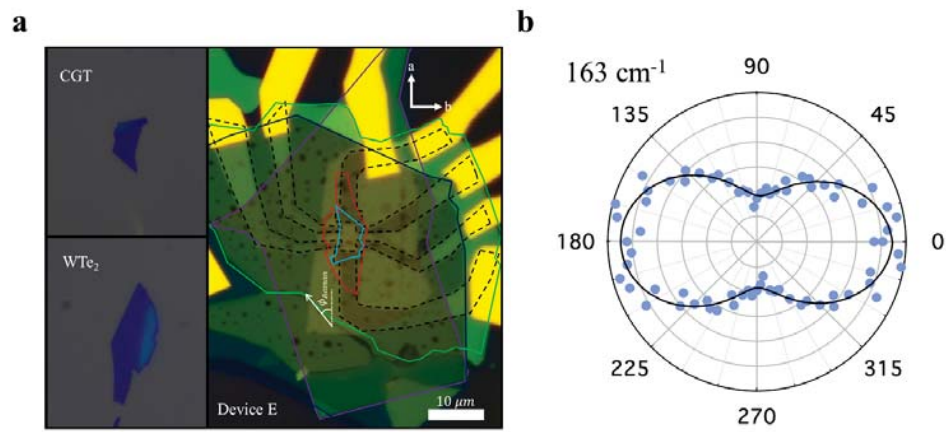

**Figure S5 | The optical images and axis characterization of device E**. **a,** the optical images of CGT and WTe$_2$ flakes (left) and the complete stack (right), where the outlines of bottom graphite (blue), bottom hBN (green), Pt (dashed black), WTe$_2$ (red), CGT (cyan), and top hBN (purple) are indicated. The axes of WTe$_2$ are labeled based on the polarized Raman results. **b,** Raman data was collected as a function of polarization angle where the $\phi_{Raman} = 0°$ is aligned to the current channel (as shown in **a**). The intensity of the Raman peak at 163 cm$^{-1}$ as a function of polarization angle is plotted as a polar plot with a $cos^2(\phi_{Raman})$ fit. The maxima around 0° confirms that the $a$-axis of WTe$_2$ is aligned along the current channel.

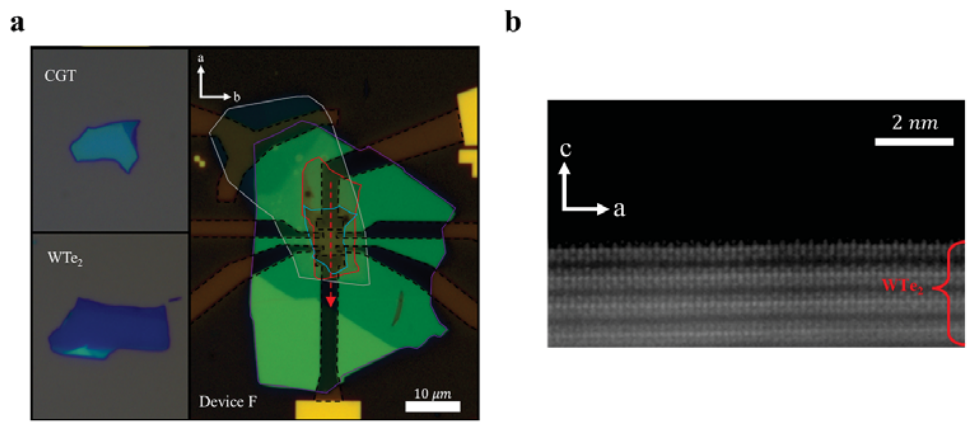

**Figure S6 | The optical images and axis characterization of device F**. **a,** the optical images of CGT and WTe$_2$ flakes (left) and the complete stack (right), where the outlines of Pt (dashed black), WTe$_2$ (red), CGT (cyan), top hBN (purple), and top graphite (grey) are indicated. The axes of WTe$_2$ are labeled based on the STEM imaging results. **b,** The cross-sectional STEM of device F taken along the red dashed arrow in **a**, showing that the cut is along the *a*-axis of WTe$_2$. We focused on the region with WTe$_2$ alone for better imaging of atomic arrangements of WTe$_2$.

**Note 3.  Temperature dependence of Longitudinal Resistance**

The longitudinal resistance ($R_{xx}$) as the function of temperature is measured for device A, B, C, D, and F as shown in Figure S7. Since we did not observe both AHE and UMR in device E, we did not perform temperature dependence measurements for device E. As shown in Figure S4a, we can measure the two-point $R_{xx}$ of WTe$_2$/ CGT bilayer and CGT separately by probing the top and bottom Hall cross region in device D. The two-point resistance of CGT alone is measured at 50 nA due to higher resistance. In Figure S7.d, we found that the resistance of CGT alone is two orders more than the resistance of WTe$_2$/ CGT bilayer, indicating that most of the current is flowing through the WTe$_2$. Note that the resistance of CGT may be underestimated here since device D is the only device that is not AFM-cleaned. Slight degradation may cause the CGT resistance to be lower. For other devices, the CGT is between the WTe$_2$ and hBN, and the electrodes are well cleaned by the AFM in the contact mode. Pristine CGT at a temperature below 100 K is expected to show insulating behavior[4]. For this reason, we can assume that the current that flows in the CGT layer in WTe$_2$/ CGT bilayer is negligible. For all the devices except device D, we estimated the resistivity by using 5 $\mu$m as the current channel width, 1.6 $\mu$m as the distance between the voltage probes, and the WTe$_2$ thickness from Table S1.

The $R_{xx}(T)$ of device A in Figure S7a shows insulating behavior, in good agreement with metal to insulator transition in WTe$_2$ when the thickness decreases to bilayer and monolayer[5]. As for device B, C, D, and F with WTe$_2$ of three or more layers, the fabricated devices all show clear metallic behavior as shown in Figure S7b-e. The metallic behavior of WTe$_2$ trilayer in device B and C also shows good interface quality, because WTe$_2$ trilayer tends to show insulating behavior due to oxidation when WTe$_2$ gets thinner[6].

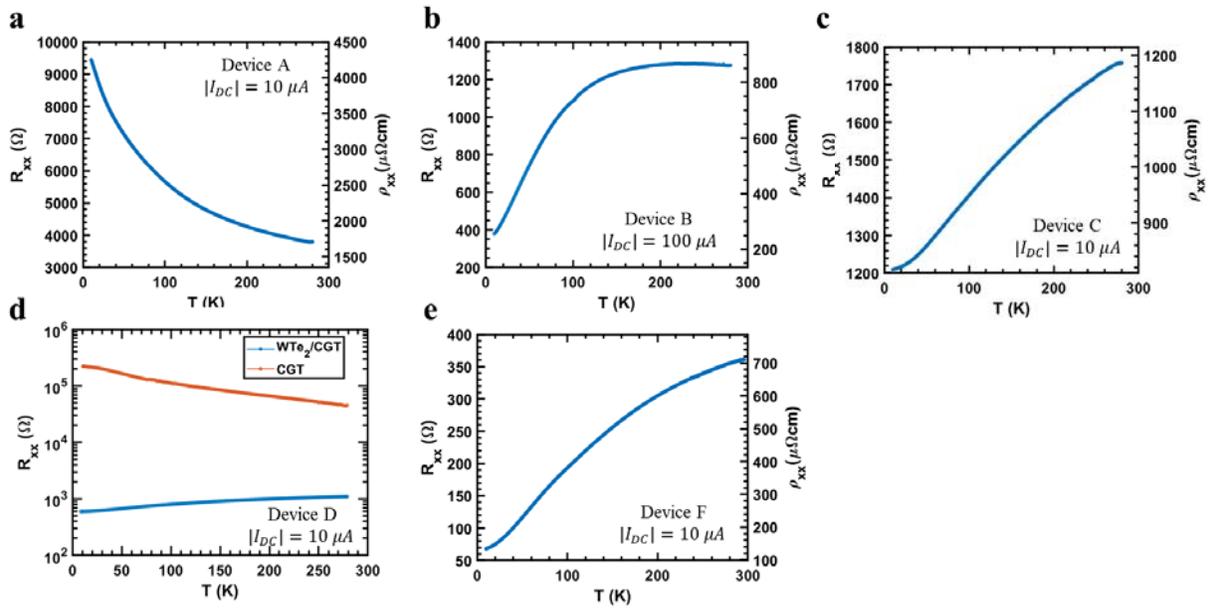

**Figure S7 | The temperature dependence of longitudinal resistance in WTe$_2$/ CGT devices.**
**a,b,c,e,** The longitudinal resistance and resistivity as a function of temperature in device A, B, C, and E. **d,** The two-point longitudinal resistance as a function of temperature in device D measured at the WTe$_2$/ CGT region and CGT region.

### Note 4. Thermal effect and control experiments

In this section, we provide some insight of thermal effect that occurs during the UMR measurements. Since the UMR typically becomes more significant when the charge current density is around $1 \times 10^{10} - 10^{11} A/m^2$, current-induced Joule heating can be pronounced. We mainly focused on device A, B, and C, where we observed unconventional UMR and performed the control experiments. To estimate the change in temperature of each device under charge current application, We utilized the temperature dependence of longitudinal resistance in supplementary note 3. As shown in Figure S8a-c, we measured the longitudinal resistance as a function of DC current magnitude in device A, B, and C. The change in longitudinal resistance when the current magnitude increases is generally due to the increase in temperatures. We used the current magnitude applied in the temperature-dependent longitudinal resistance measurements as the baseline current magnitude ($I_{base}$). For instance, we can calculate the change in longitudinal resistance at each current magnitude, defined by $\Delta R_{xx}(|I_{DC}|) = R_{xx}(|I_{DC}|) - R_{xx}(|I_{DC}| = I_{base})$, where $I_{base} = 10$ $\mu A$ in device A. At the same time we can calculate change in longitudinal resistance measured at $I_{base}$ at temperatures relative to the base temperature ($T_{base}$), where the UMR measurements were performed, defined by $\Delta R_{xx}(\Delta T) = R_{xx}(T) - R_{xx}(T = T_{base})$. We focused on the case for $T_{base} = 10K$, since that is where we performed most of the measurement and that the change in temperature due to Joule heating, given similar current level, should be comparable in the temperature range we investigated ($10 - 80K$). This way, we can obtain the change in temperature due to Joule heating as a function of $|I_{DC}|$ (Fig. S8d-f) for device A, B, and C. Due to the non-monotonic response of current-dependent resistance in device C at low current, the estimation does not work very well at lower currents. In the main text, most of the UMR measurements were performed at 100 $\mu A$, $300 \mu A$, and $200 \mu A$ in device A, B, and C, respectively. We found that the change in temperature at such current level is in general below 30 K. However, this is likely the estimated change in temperature occurred in the WTe$_2$, because the Joule heating happened in the WTe$_2$ layer, where the charge current was flowing. The change in temperature here provides an estimated upper bound for temperature increment in CGT layer due to the Joule heating.

The thermal gradient induced by Joule heating in the form of $|\nabla T| \propto I^2$ due to charge current application can lead to thermoelectric effects such as anomalous Nernst effect and spin Seebeck effect[7,8] that mimics UMR in the longitudinal resistance. The electric field due to these thermoelectric effects has the form of $\mathbf{E}_{TE} \propto \mathbf{m} \times \nabla T$, therefore, to observe change in $R_{xx}$ for perpendicular magnetization, this requires the presence of a thermal gradient in the y-direction. In a thin film device, the thermal gradient is mostly in the out-of-plane direction[8]. If there is a very small amount of in-plane thermal gradient in the y-direction, it is expected to point in opposite directions at the two opposite edges of the WTe$_2$, since the in-plane $\nabla T$ should be roughly normal to the edge of the conducting region and pointing outward. As illustrated in Figure S9a, this will result in a sign change of such magnetoresistance when different sides of voltage probes are used to measure the longitudinal resistance. We have examined the observed unconventional UMR using the voltage probes on the opposite side in device A compared to the results shown in the main text (right side in Figure S1a), while keeping the same current direction setup. The results in Figure S9b show that the measured antisymmetric part of the longitudinal resistance ($R_{xx}^{asym}$) as a function of the out-of-plane magnetic field ($B_z$) has the unconventional UMR step with similar magnitude and same sign as the one in Figure 3b in the main text. The current-dependent of UMR measured at positive and negative currents ($\Delta R_{UMR}^{+/-}$), showing similar magnitude and the same sign as Figure 3f in the main text, again confirms that the origin of this unconventional UMR should be related to the intrinsic spin-dependent electron conduction affected by the exchange coupling interactions instead of thermoelectric effects.

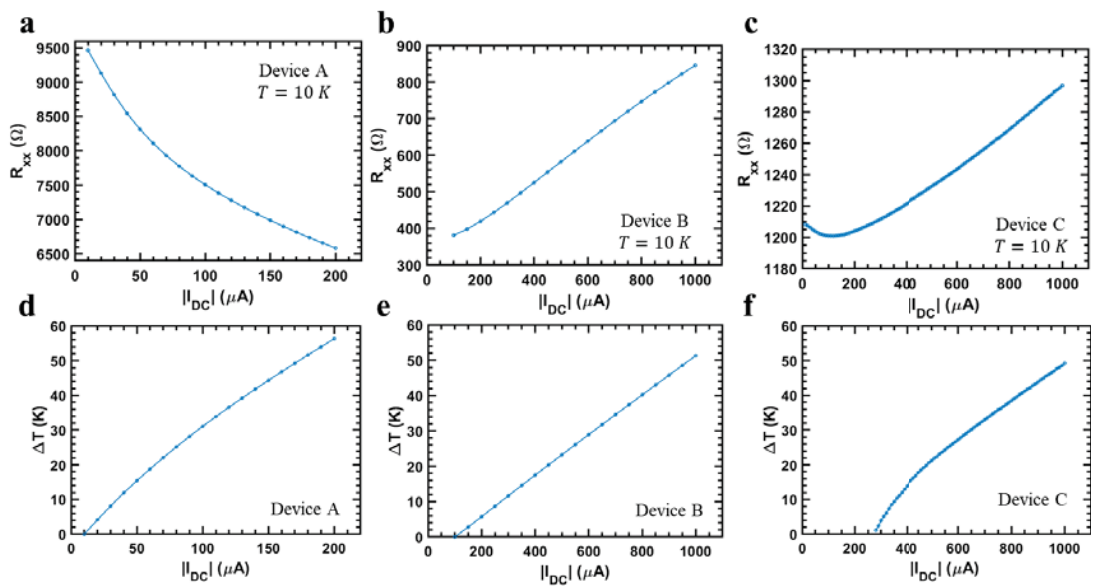

**Figure S8 | The longitudinal resistance and change in temperature ($\Delta T$) as a function of DC current magnitude**. **a,b,c,** The longitudinal resistance as a function of DC current magnitude in device A, B, and C. **d,e,f,** The estimated change in temperature as a function of DC current magnitude in device A, B, and C.

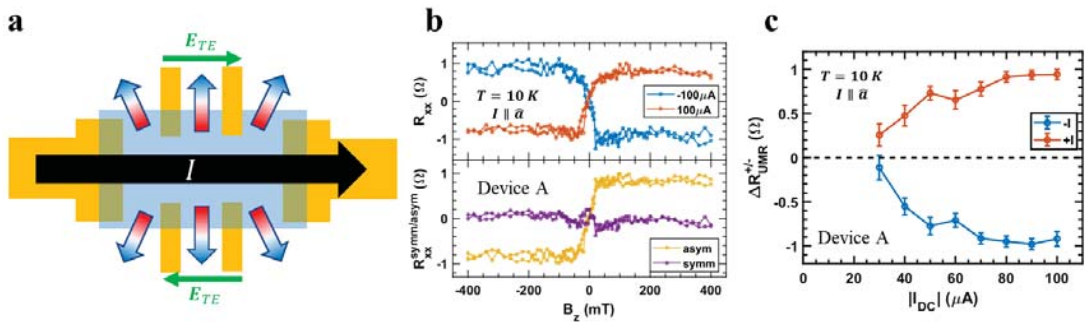

**Figure S9 | Potential thermoelectric effect and UMR results from the voltage probes on the opposite side in device A at 10 K**. **a,** A schematic showing the sign change of the electric field due to the potential thermoelectric effect corresponding to the in-plane thermal gradient that points outward from the device. **b,** Top panel: The longitudinal resistance ($R_{xx}$) as a function of out-of-plane magnetic field when a positive (orange curve) and a negative (blue curve) charge current is applied along the *a*-axis of WTe$_2$. The change of resistance due to UMR switches sign when the current is reversed. Bottom panel: The antisymmetric part ($R_{xx}^{asym}$) and the symmetric part ($R_{xx}^{symm}$) of the longitudinal resistance is plotted in yellow and purple curve, respectively. **c,** The UMR ($\Delta R_{UMR}^{(+/-)}$) as a function of the charge current magnitude $|I_{DC}|$ when a positive (orange curve) a negative (blue curve) charge current is applied along the a-axis of WTe$_2$.

## Note 5. AHE and UMR in other WTe$_2$/ CGT devices

As mentioned in supplementary note 2, in addition to device A, B, and C, we have also observed AHE in device D and F, and no UMR in device E and F. As shown in Figure S10, AHE was observed in device D when the charge current was applied along both axes of WTe$_2$. As expected, the presence of AHE does not depend on whether the current is applied along the *a* or *b*-axis of WTe$_2$. We did not try to measure UMR in device D. Note that the Anomalous Hall resistance ($\Delta R_{AHE}$) is much smaller in device D, which has a thicker WTe$_2$ (7.4 nm), in good agreement with an interface-induced effect. In device E, the transverse and longitudinal resistance as a function of the out-of-plane magnetic field (Fig. S11) measured at reasonably high current density, shows no significant sign of AHE and UMR. This may be due to the very thin CGT (1.4 nm) used in this device, resulting in reduced exchange coupling interactions. Next, we showed that in device F, which is a WTe$_2$/ CGT device with thicker WTe$_2$ (6.3 nm) aligned to the current channel in the same Hall bar geometry as device A, B, and C, the AHE can still be observed while the unconventional UMR is not observed with charge current density close to $1 \times 10^{11} A/m^2$. As shown in Figure S12a-b, the observed AHE in device F has a magnitude similar to the one observed in device D due to thicker WTe$_2$ used in the device. The antisymmetric part of the longitudinal resistance as a function of the out-of-plane magnetic field measured at various charge current magnitudes plotted in Figure S12c shows no sign of unconventional UMR, indicating that the observed UMR is an interfacial effect that only the WTe$_2$ layer in proximity with CGT contributes, as expected by the theoretical model. In addition, the unconventional UMR is due to the out-of-plane spins, which originate mainly from the WTe$_2$ layer closest to CGT, while the AHE can be contributed by either spin Hall magnetoresistance or Berry curvature induced by exchange interactions from an out-of-plane magnetization. In both cases for AHE, the magnitude of AHE can reduce when WTe$_2$ approaches the bulk limit.

Finally, the normalized UMR magnitude as a function of the charge current density in WTe$_2$/CGT devices that have the *a*-axis of WTe$_2$ parallel to the current channel in Figure S12d is a clear evidence that the UMR is negligible in the device with thicker WTe$_2$.

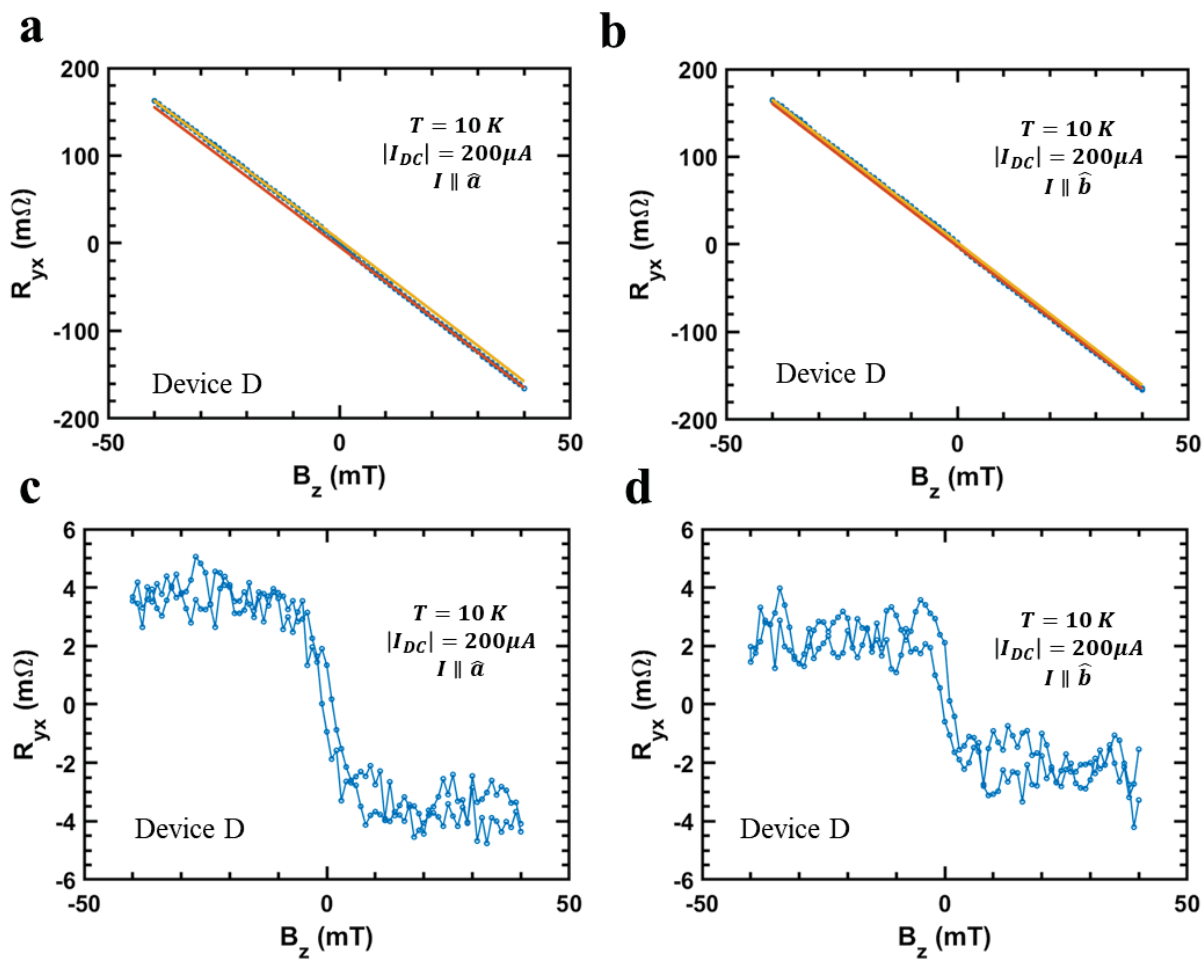

**Figure S10 | Anomalous Hall Effect in device D**. **a,b,** The raw transverse resistance as a function of the out-of-plane magnetic field with the current applied along the *a*-axis (**a**) and *b*-axis (**b**) of WTe$_2$ in device D. **c,d,** The transverse resistance as a function of the out-of-plane magnetic field with the current applied along the *a*-axis (**c**) and *b*-axis (**d**) of WTe$_2$ in device D after removing the ordinary Hall effect background.

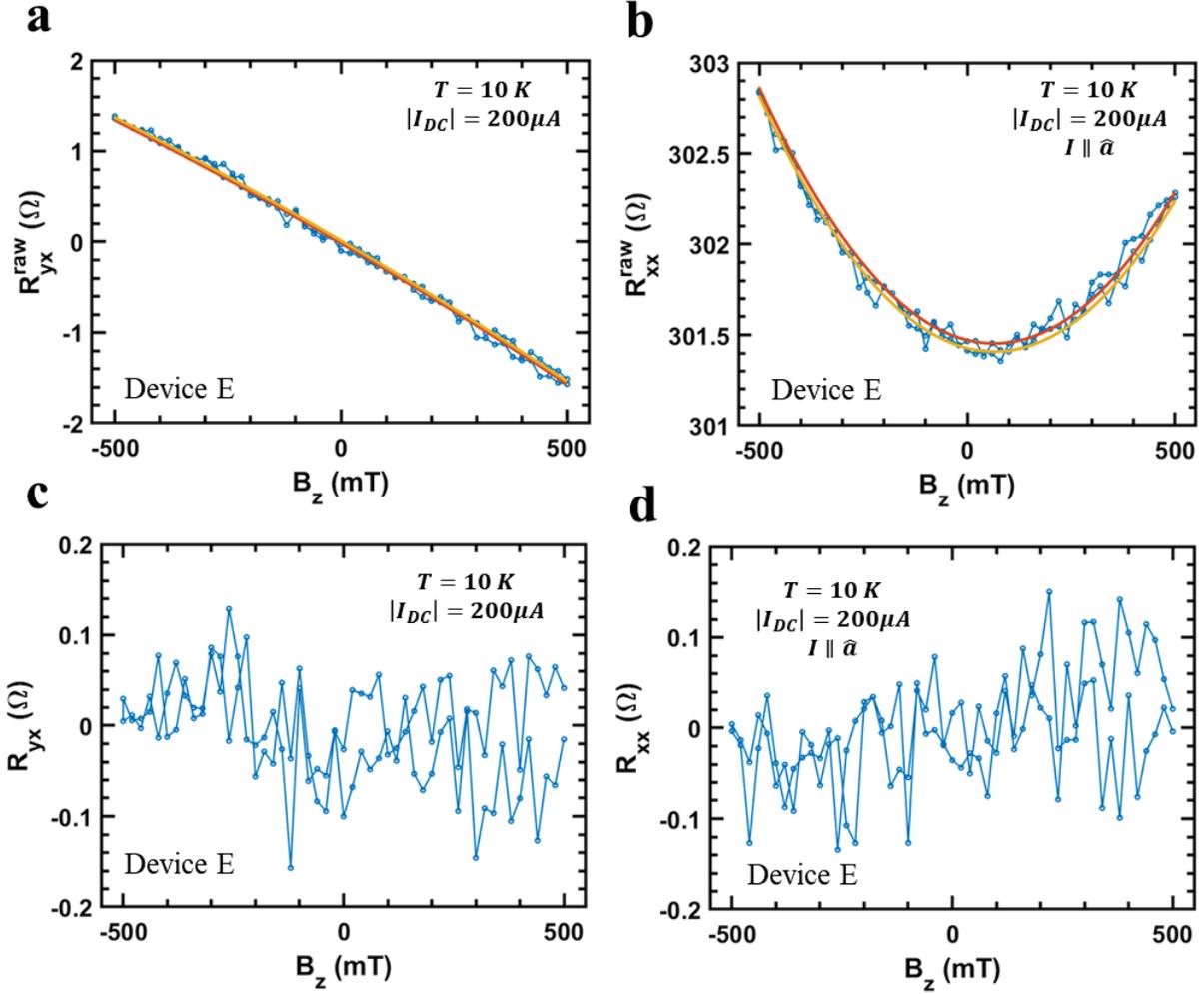

**Figure S11 | Absence of Anomalous Hall Effect and unidirectional magnetoresistance in device E. a,b,** The raw transverse (**a**) and longitudinal (**b**) resistance as a function of the out-of-plane magnetic field with the current applied along the *a*-axis of WTe$_2$ in device E. **c,d,** The transverse (**c**) and longitudinal (**d**) resistance as a function of the out-of-plane magnetic field with the current applied along the *a*-axis of WTe$_2$ in device E after removing the ordinary Hall effect and ordinary magnetoresistance background.

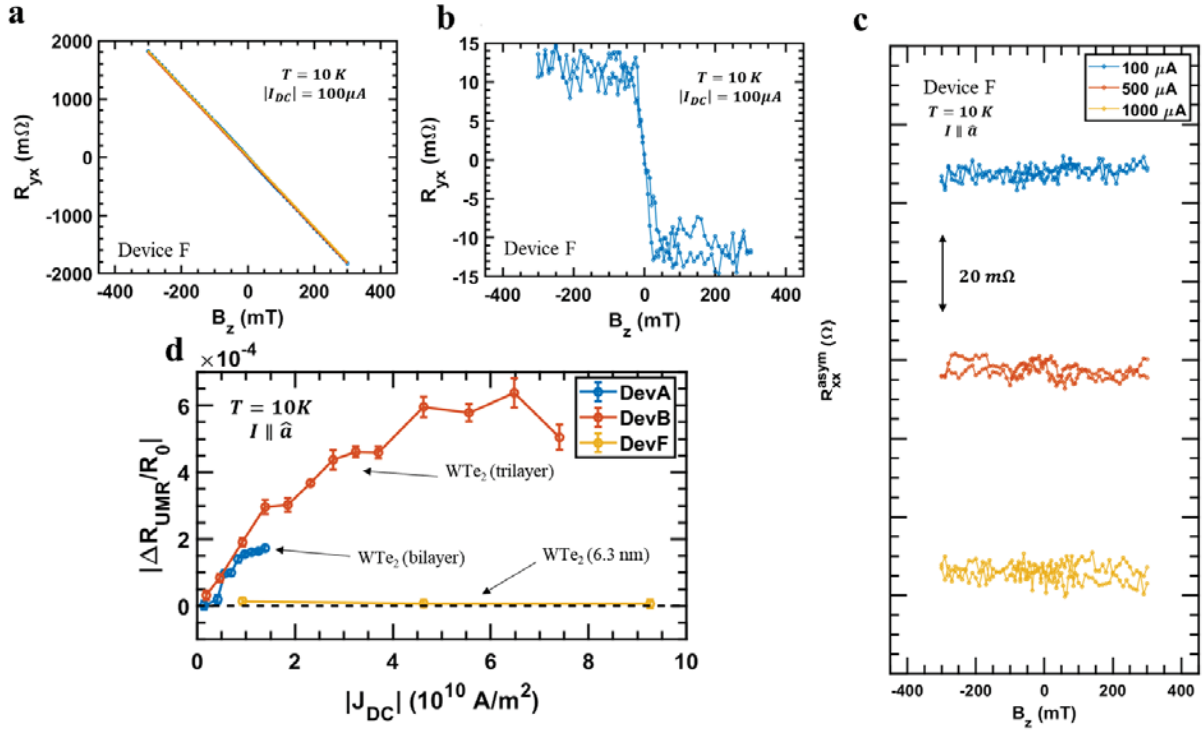

**Figure S12 | Anomalous Hall Effect and the absence of unidirectional magnetoresistance in device F**. **a,b,** The transverse resistance as a function of the out-of-plane magnetic field with the current applied along the *a*-axis of WTe$_2$ in device F before (**a**) and after (**b**) removing the ordinary Hall effect. **c,** The antisymmetric part of the longitudinal resistance as a function of the out-of-plane magnetic field at various magnitudes of current applied along the *a*-axis of WTe$_2$ in device F. **d,** The normalized UMR magnitude as a function of the charge current density measured in device A, B, and F. The unconventional unidirectional magnetoresistance is absent in device F with thicker WTe$_2$.

## Note 6. The current dependence of UMR in WTe$_2$/ CGT devices

The normalized UMR magnitude as a function of charge current density for WTe$_2$/ CGT devices in Figure 4f in the main text was obtained by measuring the $R_{xx}^{asym}(B_z)$ at various charge current magnitudes. Here, we show the $R_{xx}^{asym}(B_z)$ at some selected charge current magnitudes meausred in deivce A, B, and C in Figure S13. The results clearly show that the perpendicular magnetization sensitive unconventional UMR is only observed in devices with the current applied along the *a*-axis of WTe$_2$. Furthermore, the observed unconventional UMR became larger when the charge current magnitudes increased, indicating the quadratic-response nature of UMR.

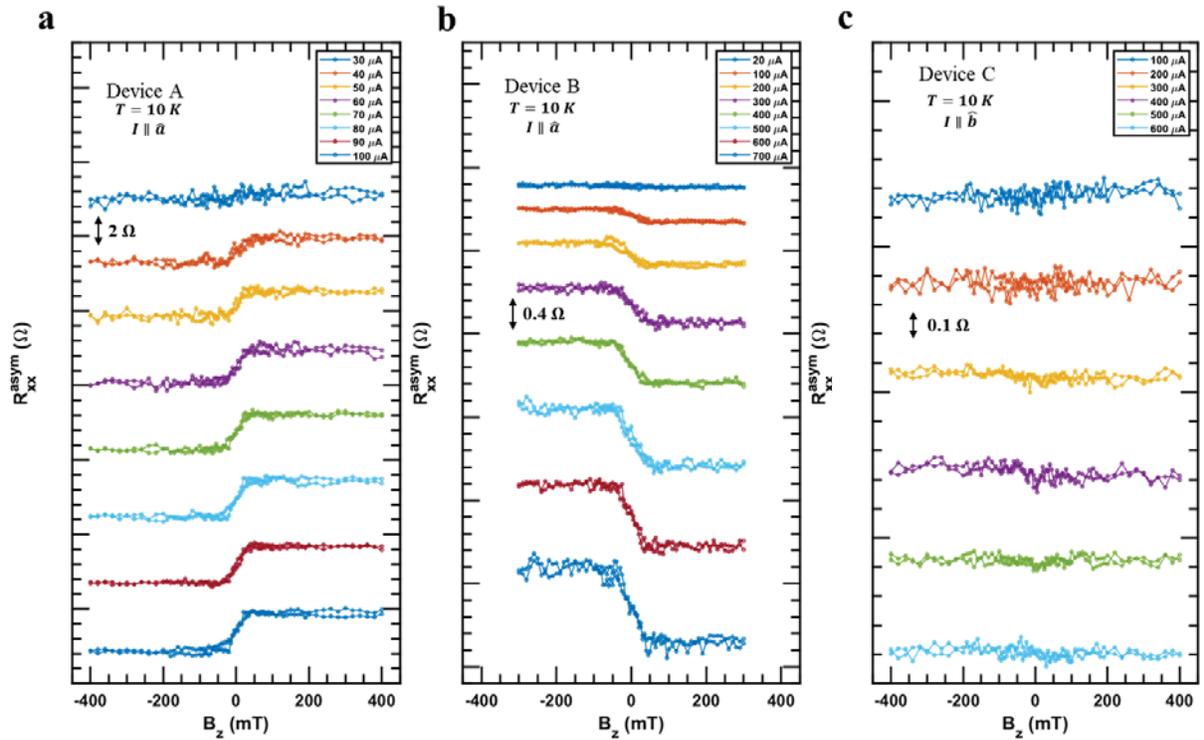

**Figure S13 | The current dependence of unconventional unidirectional magnetoresistance in WTe$_2$/ CGT devices**. **a,b,c,** The antisymmetric part of the longitudinal resistance as a function of the out-of-plane magnetic field measured at various charge current magnitudes in device A (**a**), B (**b**), and C (**c**).

**Note 7.  The temperature dependence of AHE and UMR in WTe$_2$/ CGT devices**

In this section, we present additional data for the temperature dependence of AHE and unconventional UMR observed in WTe$_2$/ CGT devices. The $R_{yx}(B_z)$ measured at temperatures ranging from 10 K to 80 K in device A, D, and F are plotted in Figure S14a-c. Figure S14d-f shows the extracted $\Delta R_{AHE}$ as the function of the temperature for device A, D, and F, where the magnitude of $\Delta R_{AHE}$ reduces when the temperature is approaching the Curie temperature (∼61 K). In general, we observed very similar temperature dependence of AHE in all WTe$_2$/ CGT devices.

The $R_{xx}^{asym}(B_z)$ measured at temperatures ranging from 10 K to 80 K in device A and B are plotted in Figure S15a-b. Figure S15c-d shows the extracted $\Delta R_{UMR}$ as the function of the temperature for device A and B, where the magnitude of $\Delta R_{UMR}$ reduces when the temperature is approaching the Curie temperature (∼61 K), similar to the temperature dependence of AHE.

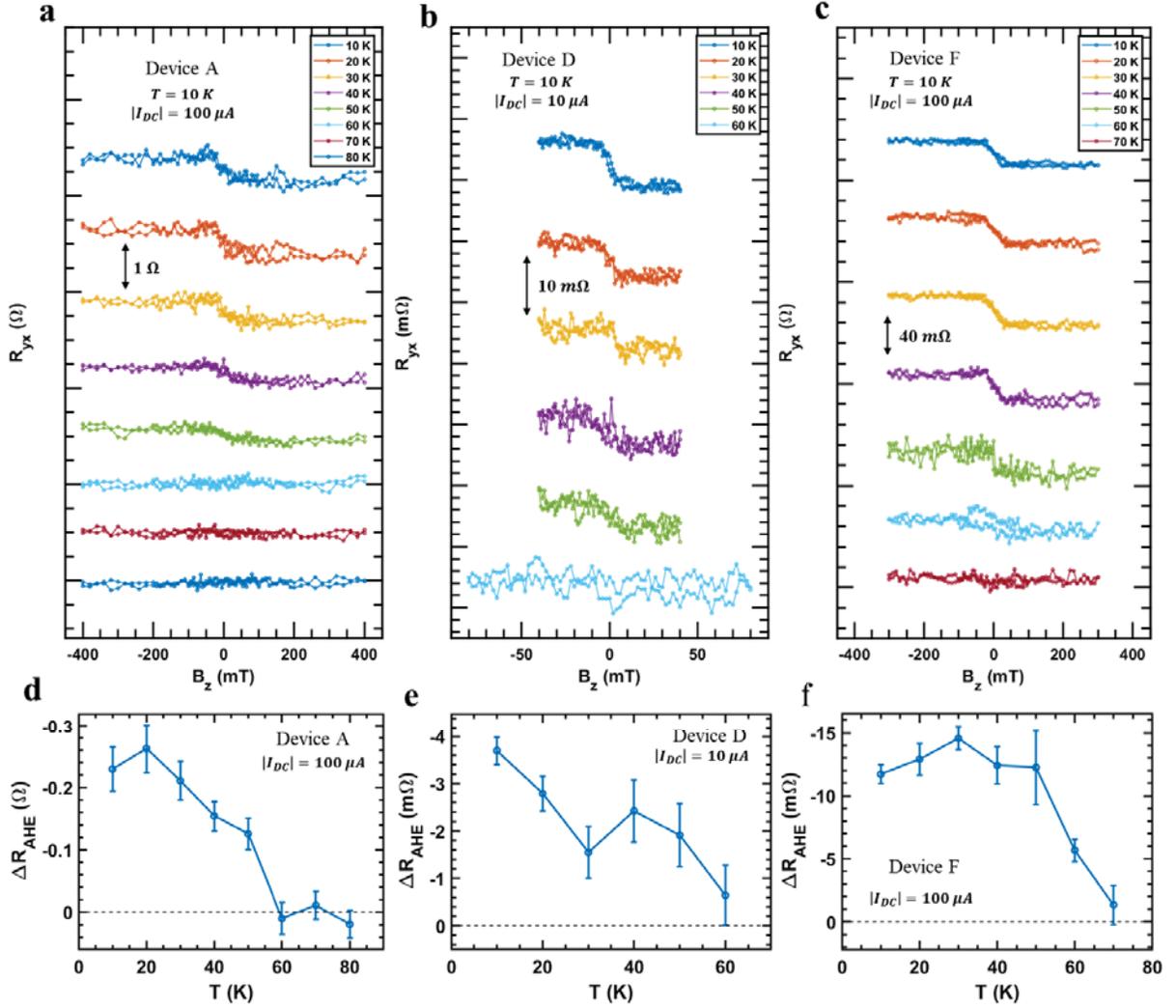

**Figure S14 | The temperature dependence of Anomalous Hall Effect in WTe$_2$/ CGT devices**. **a,b,c,** The transverse resistance as a function of the out-of-plane magnetic field measured at various temperatures in device A (**a**), D (**b**), and F (**c**). **d,e,f,** The Anomalous Hall resistance as a function of the temperature in device A (**e**), D (**e**), and F (**f**).

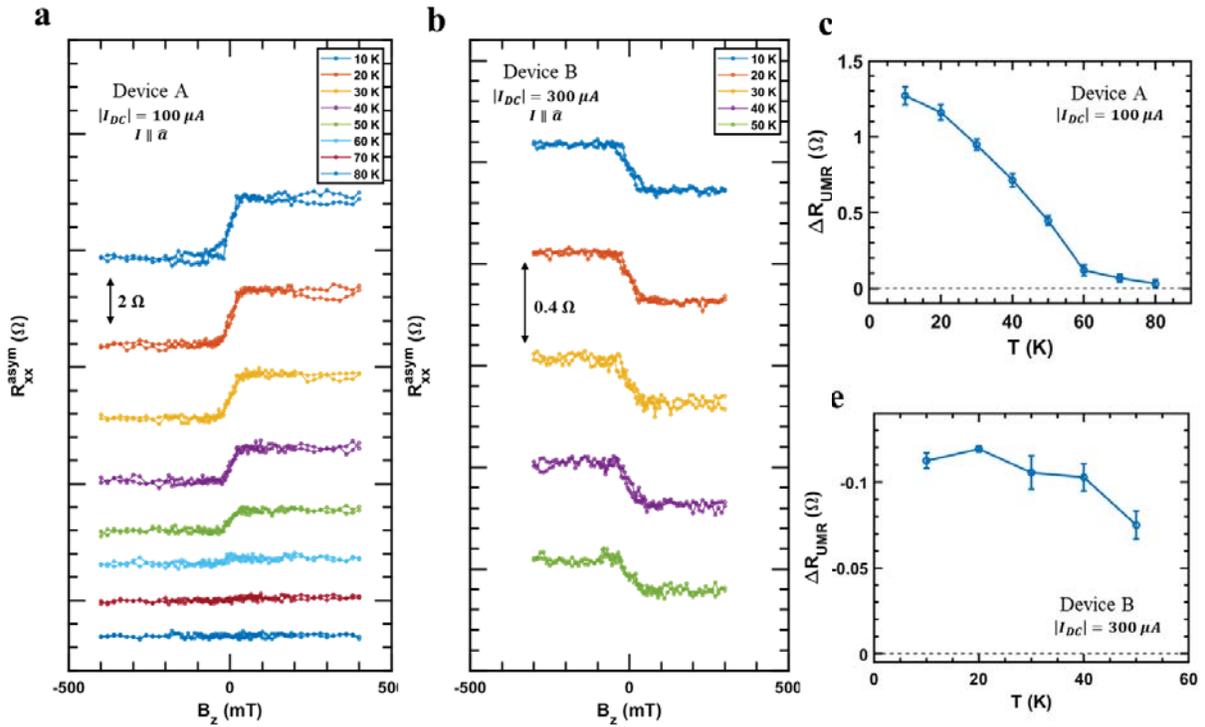

**Figure S15 | The temperature dependence of unconventional unidirectional magnetoresistance in WTe$_2$/ CGT devices**. **a,b,** The antisymmetric part of the longitudinal resistance as a function of the out-of-plane magnetic field measured at various temperatures in device A (**a**) and B (**b**). **c,d,** The unconventional unidirectional magnetoresistance as a function of the temperature in device A (**c**) and B (**d**).

## Note 8.  Effective Hamiltonian of WTe$_2$/CGT heterostructure

1T$_d$-WTe$_2$ has a mirror symmetry with respect to the $bc$-plane while the mirror symmetry is broken with respect to the $ac$-plane. Therefore, a symmetry-allowed out-of-plane non-equilibrium spin polarization can be induced by an electric field (current) along the $a$-axis. The microscopic origin of the out-of-plane spin polarization can be related to the layered-spin Edelstein effect, where only the outermost layer contributes to the net spin polarization while all the other layers partially or fully cancel each other with opposite induced spin polarization[9] due to the screw-axis and glide-plane symmetries in WTe$_2$. Additionally, the UMR in the heterostructure consisted of a ferromagnetic insulator (CGT) and a nonmagnetic conductor (WTe$_2$) has been successfully quantified using the only interfacial transport theory[10]. We also have observed a negligible UMR signal in the bulk WTe$_2$ sample which justifies the fact that the UMR originates from the interfacial current which will be overwhelmed by the bulk current in bulk WTe$_2$.

Therefore, to characterize the UMR in WTe$_2$/CGT heterostructure, we can model such heterostructure by introducing two additional terms, i.e., exchange coupling interaction and interfacial Rashba spin-orbit coupling, into the 1T$_d$-monolayer WTe$_2$ Hamiltonian[11] to account for the interactions from the magnetization and symmetry breaking at the interface. The Hamiltonian reads

$$H = \epsilon_0(k_x, k_y) + \beta \sin(k_y a_y)\sigma_0 \otimes \tau_0 + \eta \sigma_0 \otimes \tau_x + H_{soc} + H_{ex} + H_R, \qquad (1)$$

where the hopping term is $\epsilon_0 = m_p[4-2\cos(k_x a_x)-2\cos(k_y a_y)]\sigma_0 \otimes \tau_0 + m_d[4-2cos(k_x a_x) - 2cos(k_y a_y) + \delta]\sigma_0 \otimes \tau_z$ with $\boldsymbol{\sigma}$ and $\boldsymbol{\tau}$ the Pauli matrices for the spin and sublattice degree of freedom, respectively. The $\beta$ and $\eta$ terms represent the $x$-$y$ crystalline anisotropy and the broken inversion symmetry respectively. The intrinsic SOC interaction is given by $H_{soc} = \Lambda_x \sin(k_y a_y)\sigma_x \otimes \tau_x + \Lambda_y \sin(k_x a_x)\sigma_y \otimes \tau_x + \Lambda_z \sin(k_x a_x)\sigma_z \otimes \tau_x$. The interfacial Rashba SOC and exchange couplings between the electrons in WTe$_2$ and the magnetization of CGT are given by $H_R = -\alpha_R[\sin(k_y a_y)\sigma_x \otimes \tau_0 - \sin(k_x a_x)\sigma_y \otimes \tau_0]$ and $H_{ex} = -\Delta_{ex}\sigma_z \otimes \tau_0$, respectively. In our convention, we set the $a$-axis of WTe$_2$ to be the $\hat{x}$ direction and the $b$-axis to be the $\hat{y}$ direction. The values of the parameters adopted in our numerical calculations are detailed in Table S2. They are obtained by fitting the band structure and spin texture of the effective Hamiltonian of 1T$_d$-monolayer WTe$_2$ with the DFT results[11].

**Table S2** | Values of the Hamiltonian parameters (in unit of meV) adopted in the numerical calculation.

| $(m_p, m_d)$ | $\delta$ | $\beta$ | $\eta$ | $(\Lambda_x, \Lambda_y, \Lambda_z)$ |
|---|---|---|---|---|
| (-105, -544.9) | 424.8 | 449.4 | 1.7 | (59.1, 77.7, -115.9) |

We then calculate the electronic band structures of our effective Hamiltonian (Eq. 1). The first and second conduction bands are shown in the Figure S16. One can see that the exchange interaction lifts the spin degeneracy in the nonmagnetic WTe$_2$ while the Rashba SOC introduces the mixing of the spin. Additionally, it can be seen from the inset figures that the exchange interaction from the adjacent CGT layer introduces the asymmetry in the band structure. This assymetry can be further enlarged by the Rashba SOC (Fig. S16a). Note that, the asymmetry exists even without the Rashba SOC (Fig. S16c). This asymmetry induced by the out-of-plane magnetization is related to the out-of-plane spin-Edelstein effect (Fig. S20) and is responsible for the UMR in the electronic transport.

The Berry curvature of our effective Hamiltonian (Eq. 1) is given by $\Omega^n = \nabla_{\bm{k}} \times \mathcal{A}^n_{\bm{k}}$ with Berry connection for $n$-th band defined as $\mathcal{A}^n_{\bm{k}} = i\langle n|\partial_{\bm{k}}|n\rangle$. The Berry curvature for first and second conduction and valence bands are shown in Figure S17. With $\Delta_{ex}$=50 meV, and $\alpha_R$=10 meV, we find that the topological Chern numbers are 1 for the first conduction and valence bands and -1 for the second conduction and valence bands.

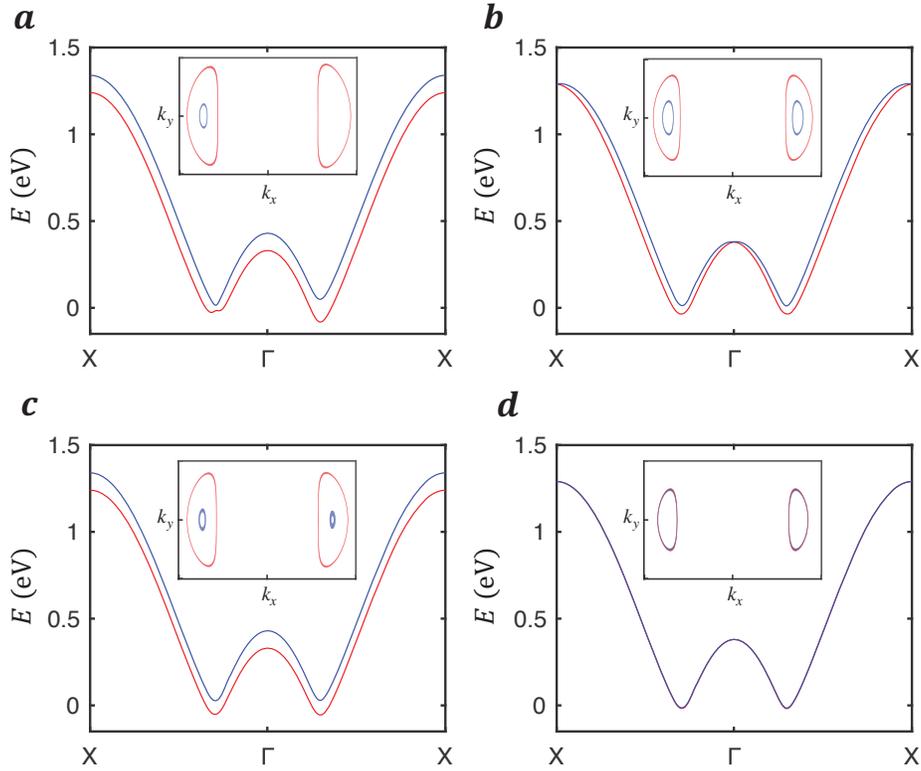

**Figure S16 | Band structure along the high-symmetry line** $\mathrm{X}-\Gamma-\mathrm{X}$ **of WTe$_2$/CGT with different exchange and Rashba interactions strength.**. **(a)**: $\Delta_{ex}$=50 meV, $\alpha_R$=50 meV. **(b)**: $\Delta_{ex}$=0 meV, $\alpha_R$=50 meV. **(c)**: $\Delta_{ex}$=50 meV, $\alpha_R$=0 meV. **(d)**: $\Delta_{ex}$=0 meV, $\alpha_R$=0 meV. For the clarity purpose, we adapt a large $\alpha_R = \Delta_{ex}$ to enlarger the asymmetry in the band plotting here. In numerical calculation of UMR, $\alpha_R$ is set to be 10 meV. Inset figures: The corresponding 2D band cut in the $k_x$-$k_y$ plane for energy around 30 meV.

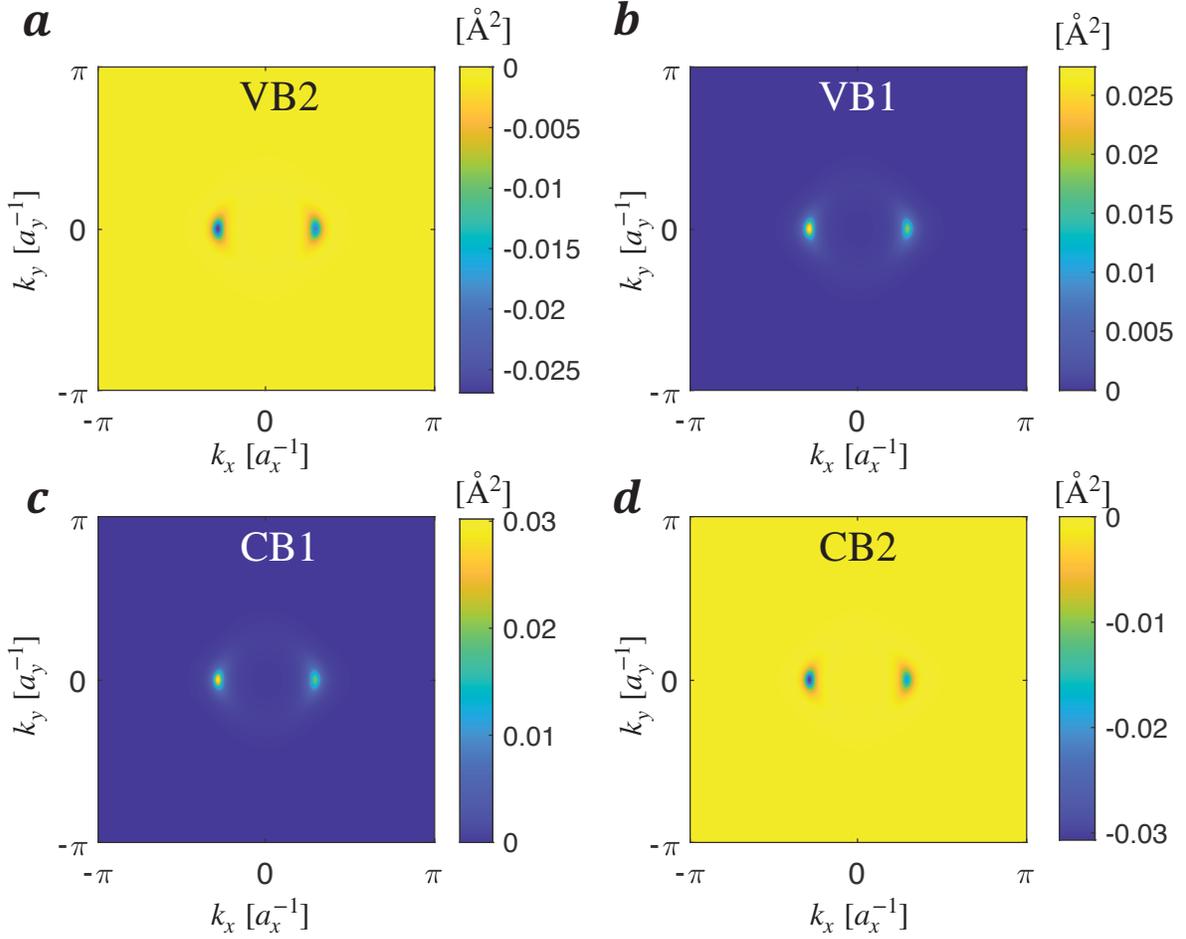

**Figure S17 | K-space distribution of Berry curvature**. **a-d,** Berry curvature of first and second valence bands (VB) and conduction bands (CB). Parameters: $\Delta_{ex}$=50 meV, $\alpha_R$=10 meV.

**Note 9. Theory of unidirectional magnetoresistance in WTe$_2$/CGT**

The UMR characterizes the asymmetry in the longitudinal resistance when the direction of the applied electric field (or magnetization) is reversed. This results in distinguishable positive and negative directional responses, rendering them unidirectional. We first assume an electric field is applied in the $x$ direction, which can be straightforwardly generalized to the y direction. With the longitudinal resistivity defined as $\rho_{xx}(E_x) = E_x/J_x(E_x)$, the UMR ratio can be written as

$$\text{UMR}(E_x) = \frac{\rho_{xx}(E_x) - \rho_{xx}(-E_x)}{\rho_{xx}(E_x) + \rho_{xx}(-E_x)} = -\frac{j_x(E_x) + j_x(-E_x)}{j_x(E_x) - j_x(-E_x)} \tag{2}$$

If there is no asymmetry in the charge current for $\pm E_x$ field, i.e., $j_x(-E_x) = -j_x(E_x)$, then the UMR vanishes. To the second order of electric field, the longitudinal current has the following expansion form $j_x = j_x^{(1)} + j_x^{(2)} = \sigma_{xx}^{(1)} E_x + \sigma_{xx}^{(2)} E_x^2 + \cdots$. The second-order conductivity $\sigma_{xx}^{(2)}$ introduces a nonlinear charge current $j_x^{(2)}$ that is even with respect to the applied electric field. Therefore, it remains invariant after reversing the electric field, thus causing the asymmetry in the longitudinal response that is responsible for the UMR. This nonlinear charge current can be understood as a result of the partial conversion of a second-order spin current[12]. The UMR defined in Eq. 2 is approximately equal to the normalized UMR obtained from experiments, i.e., $\Delta R_{UMR}/R_0$, since the second-order response is much smaller than the first-order response ($j_x^{(1)} \gg j_x^{(2)}$), meaning that reversing the electric field is close to reversing the current polarity. By inserting $j_x$ into Eq. 2, we obtain a simple expression for the UMR that is proportional to the second-order conductivity and the inverse of first-order conductivity: $\text{UMR} = -\frac{\sigma_{xx}^{(2)}}{\sigma_{xx}^{(1)}} E_x$.

The extrinsic Fermi surface contribution for the longitudinal conductivity of first and second order are given by:

$$\sigma_{xx}^{(1)} = e^2 \tau \sum_n \int_{\text{BZ}} \frac{dk^2}{(2\pi)^2} (\hat{v}_x^{nn})^2 \frac{2\Gamma^3}{\pi[(\epsilon_n - \mu)^2 + \Gamma^2]^2}, \tag{3}$$

$$\sigma_{xx}^{(2)} = -\frac{e^3 \tau^2}{\hbar} \sum_n \int_{\text{BZ}} \frac{dk^2}{(2\pi)^2} \hat{v}_x^{nn} (\partial_{k_x} \hat{v}_x)^{nn} \frac{2\Gamma^3}{\pi[(\epsilon_n - \mu)^2 + \Gamma^2]^2}, \tag{4}$$

where the relaxation time is $\tau = \hbar/2\Gamma$ with $\Gamma$ the disorder broadening and velocity operator $\hbar v_x = \partial_{k_x} \hat{H}/\hbar$. Its matrix element is defined as $\hat{v}_x^{nn} = \langle n | \hat{v}_x | n \rangle$. The summation is for all the energy bands $|n\rangle$ and $2\Gamma^3/\pi[(\epsilon_n - \mu)^2 + \Gamma^2]^2$ is a Lorentzian function with a broadening $\Gamma$ and a center at the chemical potential $\mu_F$. Therefore, only the bands that are near the Fermi surface contribute to the longitudinal conductivity.

On the other hand, the anomalous Hall current $j_x^{(1)} = \sigma_{xy} E_y$ can be characterized by the first-order transverse conductivity $\sigma_{xy}^{(1)}$, which can be calculated from the intrinsic Fermi sea contribution:

$$\sigma_{xy}^{(1)} = -2e^2 \hbar \sum_n f(\epsilon_n, \mu_F) \sum_{m \neq n} \int \frac{dk^2}{(2\pi)^2} \text{Im}[\hat{v}_y^{nm} \hat{v}_x^{mn}] \frac{(\epsilon_n - \epsilon_m)^2 - \Gamma^2}{[(\epsilon_n - \epsilon_m)^2 + \Gamma^2]^2}, \tag{5}$$

where $f(\epsilon_n, \mu_F)$ is the Fermi-Dirac distribution function. Therefore, all the occupied bands below the Fermi surface contribute to the anomalous Hall conductivity. Note that the skew scattering and side jump mechanisms, which belong to the extrinsic contribution, are not included in the above formalism with constant relaxation time approximation.

In Figure S18a and Figure S18b, we plot the calculated longitudinal and transverse conductivity of first and second order as a function of Rashba SOC strength. The dependence on exchange coupling strength $\Delta_{ex}$ can be found in the main text. The calculated UMR for different directions of an applied electric field is also illustrated in Figure S18c and Figure S18d. We find that the first-order longitudinal conductivity (extrinsic) is much larger than the transverse conductivity (intrinsic) as expected in the clean limit (small $\Gamma$). Additionally, we find that in the second-order longitudinal conductivity, the $\sigma_{xx}^{(2)}$ is finite even without the Rashba SOC while the $\sigma_{yy}^{(2)}$ is always vanishing. This results in a finite UMR in the $\hat{x}$ direction and vanishing UMR in the $\hat{y}$ direction (Fig. S18c and Fig. S18d). Note that the Rashba SOC originated from the interface symmetry breaking can increase the second-order conductivity and thus the UMR ratio.

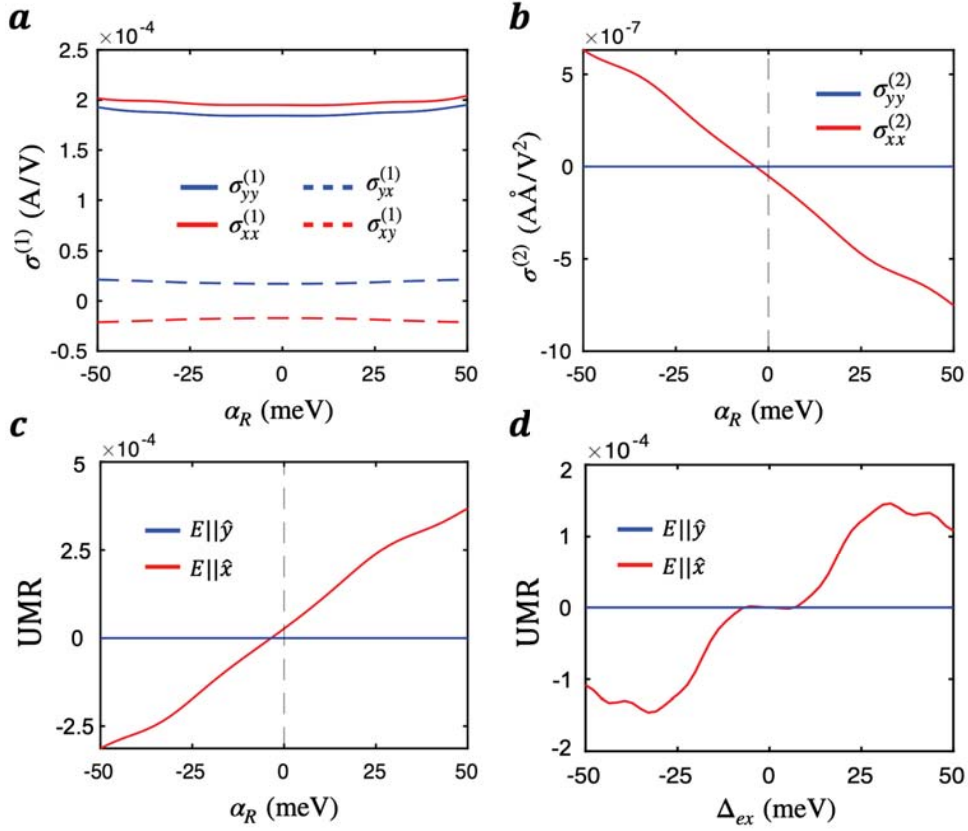

**Figure S18 | Longitudinal and transverse conductivity of first and second order and UMR**. **(a)** First-order conductivity and **(b)** second-order conductivity as a function of Rashba SOC. **(c)** The calculated UMR as a function of Rashba SOC and **(d)** exchange coupling strength $\Delta_{ex}$. Parameters: $\Delta_{ex}$=50 meV, $\alpha_R$=10 meV, $|E| = 10^5$ V/m.

In Figure S19, we also plot the dependence on the chemical potential $\mu_F$ for the first-and second-order conductivity, where we find that when the $\mu_F$ approaches the edges of the valence and conduction bands, the first-order transverse (AHE) conductivity $\sigma_{xy}^{(1)}$ and $\sigma_{yx}^{(1)}$ will decrease while the second-order longitudinal conductivity $\sigma_{xx}^{(2)}$ and $\sigma_{yy}^{(2)}$ will increase. The decreasing AHE conductivity near the edge of conduction bands (valence bands) is consistent with the opposite Chern numbers of the two conduction bands (valence bands). Thus, when $\mu_F$ lies inside the gap, the AHE conductivity becomes vanishing (no quantized AHE). As expected, the first-order longitudinal conductivity $\sigma_{xx}^{(1)}$ and $\sigma_{yy}^{(1)}$ will increase when $\mu_F$ is set away from the gap.

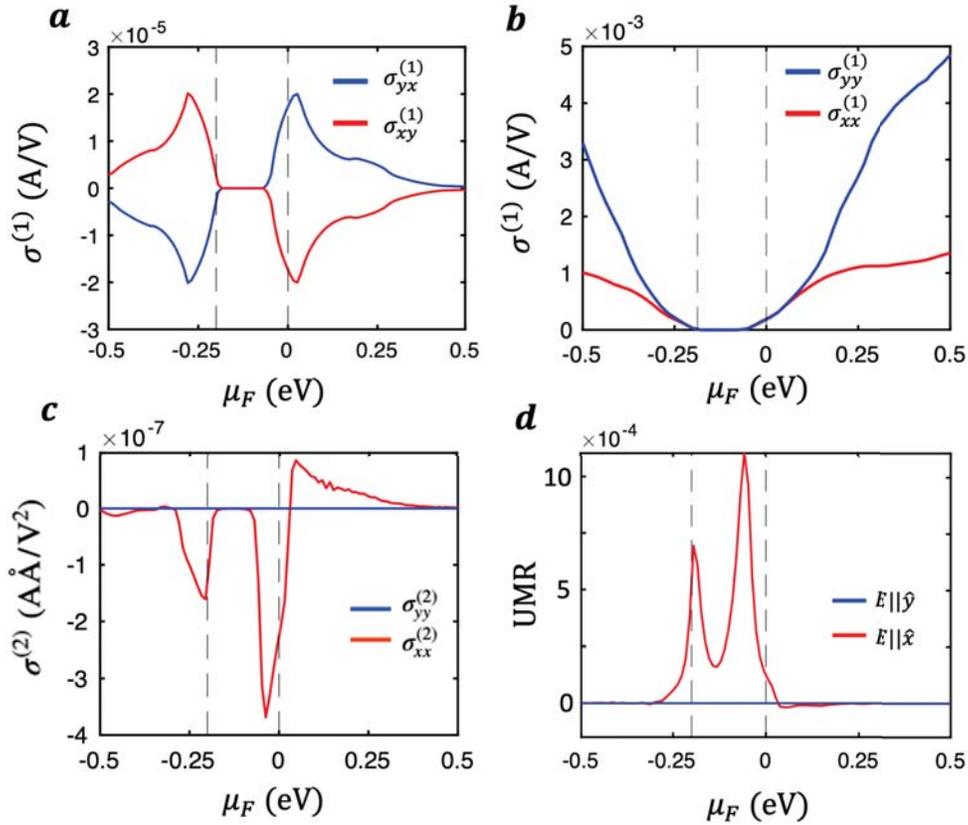

**Figure S19 | The dependence on chemical potential for conductivity and UMR. a,b,** First-order transverse (**a**) and longitudinal (**b**) conductivity as a function $\mu_F$. (**c**) The second-order longitudinal conductivity and (**d**) UMR ratio as a function of $\mu_f$. The two dashed lines locate at the valence band edge and $E_f = 0$ (default value, near the conduction band edge), respectively. Parameters: $\Delta_{ex}$=50 meV, $\alpha_R$=10 meV, $|E| = 10^5$ V/m.

Note that in 2D, the current density is defined as the current per length $j_x^{2D} = I_x/w$ with $w$

the sample width. Therefore, the transverse resistance and resistivity are the same $\rho_{xy} = R_{xy}$. While the longitudinal resistance and resistivity differ only by a dimensionless geometry factor $R_{xx} = \rho_{xx} L/w$ with $L$ the sample length. The first-order conductivity and conductance also have the same unit in 2D. With the definition $\sigma^{(1)} = j^{2D}/E$, $\sigma^{(1)}$ has the unit of A/V, which is the same with the unit of $e^2/h$. With this convention, the second-order conductance has the unit of m · A/V². We can estimate that a charge current density of $J_{DC} = 1 \times 10^{10}$ $A/m^2$ corresponds to an applied electric field $E \sim 10^5$ V/m. With such electric field applied along the $a$-axis, the UMR= $-\frac{\sigma^{(2)}_{xx}}{\sigma^{(1)}_{xx}} E_x$ is estimated to be at the order of $10^{-4}$, which matches well the experimental results. Unless otherwise specified, the default parameters of exchange coupling strength, Rashba SOC, Fermi level and disorder broadening in our numerical calculations are $\Delta_{ex} = 50$ meV, $\alpha_R = 10$ meV, $E_f = 0$, $\Gamma = 10$ meV with a $400 \times 200$ mesh grid for the sampling in the first Brillouin zone.

## Note 10. Spin-Edelstein effect in WTe$_2$/CGT

To understand why the UMR is present in the $\hat{x}$ direction and absent in the $\hat{y}$ direction, we turn to the non-equilibrium spin density $\boldsymbol{\delta S} = \chi \boldsymbol{E}$ induced by the applied electric field $\boldsymbol{E}$ in the linear response theory (spin-Edelstein effect). Note that 1T$_d$-WTe$_2$ breaks the mirror symmetry along the $b$-axis while preserves the mirror symmetry along $a$-axis ($bc$-mirror plane), which results in a nonvanishing $\delta S_z$ induced by an electric field along the $a$-axis ($\hat{x}$ direction). On the other hand, if the electric field is applied along the $b$-axis ($\hat{y}$ direction), the $\delta S_z$ should be vanishing as required by the symmetry[9]. To justify this, we use the Kubo formula to calculate the non-equilibrium spin density, which is given by

$$\delta S_\beta = \frac{e\hbar}{2\Gamma} \sum_n \int_{\text{BZ}} \frac{dk^2}{(2\pi)^2} \hat{S}_\beta^{nn} \hat{v}_\alpha^{nn} E_\alpha \frac{2\Gamma^3}{\pi[(\epsilon_n - \mu_F)^2 + \Gamma^2]^2}, \tag{6}$$

where the spin operator is given by $\hat{S}_\beta = \sigma_\beta \otimes \tau_0$. In Figure S20, we plot the non-equilibrium spin density $\boldsymbol{\delta S}$ (per nanometer) with respect to the Rashba SOC $\alpha_R$ and exchange coupling strength $\Delta_{ex}$ for different directions of the applied electric field. We find our numerical results well match the conclusion predicted by the symmetry analysis: For electric field along the $\hat{x}$ direction ($a$-axis), only the $\delta S_x$ is vanishing while both $\delta S_y$ and $\delta S_z$ can be nonvanishing. On the other hand, for the electric field along the $\hat{y}$ direction ($b$-axis), only the $\delta S_x$ can be nonvanishing. Consequently, reversing the electric field in the $a$-axis reverses the sign of $\delta S_z$, causing the asymmetry in the electronic transport for a fixed magnetization in the $\hat{z}$ direction ($c$-axis). Equivalently, reversing the magnetization in the $\hat{z}$ direction ($c$-axis) will cause the asymmetry in the electronic transport for a fixed direction of $\boldsymbol{E}$ field.

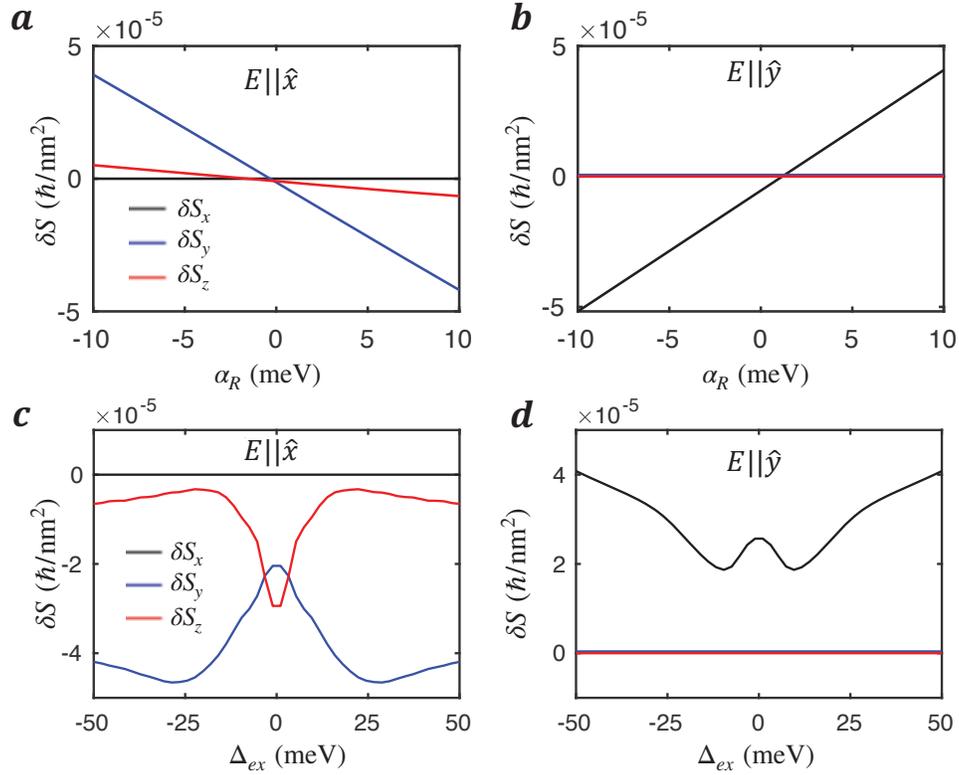

**Figure S20 | Spin-Edelstein effect in WTe$_2$/CGT. a,b,** $\delta S_x$, $\delta S_y$, $\delta S_z$ as a function of Rashba SOC strength $\alpha_R$ for electric field along $\hat{x}$ (**a**) and $\hat{y}$ (**b**) direction. **c,d,** $\delta S_x$, $\delta S_y$, $\delta S_z$ as a function of exchange coupling strength $\Delta_{ex}$ for electric field along $\hat{x}$ (**c**) and $\hat{y}$ (**d**) direction. non-equilibrium spin density $\boldsymbol{\delta S}$ in unit of $\hbar$ per nanometer. Parameters: $\Delta_{ex}$=50 meV, $\alpha_R$=10 meV, $E_f$=0, $\Gamma$=10 meV, $|E| = 1\text{V}/\mu\text{m}$.